\documentclass[12pt,a4paper]{article}
\usepackage{graphicx,color}
\usepackage{amsmath}
\usepackage{amssymb}
\pdfoutput=1
\usepackage{multirow}
\usepackage{url}
\usepackage{subfigure}
\usepackage{authblk}
\usepackage{fullpage}
\usepackage{lscape}
\begin{document}

\thispagestyle{empty}
\title{\Huge{Proposal: JSNS$^2$-II}}

\author[16]{\small{S.~Ajimura}} 
\author[15]{M.~Botran}
\author[4]{J.~H.~Choi}
\author[3]{J,~W.~Choi}
\author[19]{M.~K.~Cheoun}
\author[22]{T.~Dodo}
\author[22]{H.~Furuta}
\author[13]{J.~Goh}
\author[7]{K.~Haga}
\author[7]{M.~Harada} 
\author[7,8]{S.~Hasegawa} 
\author[22]{Y.~Hino}
\author[16]{T.~Hiraiwa} 
\author[17]{H.~I.~Jang}
\author[6]{J.~S.~Jang}
\author[3]{M.~C.~Jang}
\author[20]{H.~Jeon}
\author[20]{S.~Jeon}
\author[3]{K.~K.~Joo}
\author[15]{J.~R.~Jordan} 
\author[20]{D.~E~Jung} 
\author[18]{S.~K.~Kang}
\author[7]{Y.~Kasugai} 
\author[11]{T.~Kawasaki} 
\author[9]{E.~J.~Kim}
\author[3]{J.~Y.~Kim}
\author[20]{S.~B.~Kim\footnote{Co-spokesperson}}
\author[15]{S.~Y.~Kim}
\author[14]{W.~Kim}
\author[11]{T.~Konno}
\author[7]{H.~Kinoshita}
\author[10]{D.~H.~Lee}
\author[13]{S.~Lee}
\author[3]{I.~T.~Lim}
\author[15]{E.~Marzec} 
\author[10]{T.~Maruyama\footnote{Spokesperson}}
\author[7]{S.~Masuda}
\author[7]{S.~Meigo} 
\author[10]{S.~Monjushiro}
\author[3]{D.~H.~Moon}
\author[16]{T.~Nakano} 
\author[12]{M.~Niiyama}
\author[10]{K.~Nishikawa{\footnote{Deceased}}}
\author[16]{M.~Nomachi} 
\author[4]{M.~Y.~Pac}
\author[10]{J. S. Park}
\author[21]{S.~J.~M.~Peeters}
\author[5]{H.~Ray}
\author[20]{G.~Roellinghoff}
\author[20]{C.~Rott} 
\author[7]{K.~Sakai} 
\author[7]{S.~Sakamoto}
\author[16]{T.~Shima} 
\author[3]{C.~D.~Shin}
\author[15]{J.~Spitz} 
\author[1]{I.~Stancu}
\author[16]{Y.~Sugaya}
\author[22]{F.~Suekane}
\author[7]{K.~Suzuya}
\author[10]{M.~Taira}
\author[22]{R.~Ujiie}
\author[7]{Y.~Yamaguchi}
\author[2]{M.~Yeh}
\author[17]{I.~S.~Yeo}
\author[13]{C.~Yoo}
\author[20]{I.~Yu} 
\author[3]{A.~Zohaib}

\affil[1]{\scriptsize{University of Alabama, Tuscaloosa, AL, 35487, USA}}
\affil[2]{Brookhaven National Laboratory, Upton, NY, 11973-5000, USA}
\affil[3]{Department of Physics, Chonnam National University, Gwangju, 61186, KOREA}
\affil[4]{Laboratory for High Energy Physics, Dongshin University, Chonnam 58245, KOREA}
\affil[5]{University of Florida, Gainesville, FL, 32611, USA}
\affil[6]{Gwangju Institute of Science and Technology, Gwangju, 61005, KOREA}
\affil[7]{J-PARC Center, JAEA, Tokai, Naka Ibaraki 319-1195, JAPAN}
\affil[8]{Advanced Science Research Center, JAEA, Tokai, Naka Ibaraki 319-1195, JAPAN}
\affil[9]{Division of Science Education, Jeonbuk National University, Jeonju, 54896, KOREA}
\affil[10]{High Energy Accelerator Research Organization (KEK), Tsukuba, Ibaraki 305-0801, JAPAN}
\affil[11]{Department of Physics, Kitasato University, Sagamihara, Kanagawa 252-0373, JAPAN}
\affil[12]{Department of Physics, Kyoto Sangyo University, Kyoto 603-8555, JAPAN}
\affil[13]{Department of Physics, Kyung Hee University, Seoul 02447, Korea}
\affil[14]{Department of Physics, Kyungpook National University, Daegu 41566, KOREA}
\affil[15]{University of Michigan, Ann Arbor, MI, 48109, USA}
\affil[16]{Research Center for Nuclear Physics, Osaka University, Osaka 565-0871, JAPAN}
\affil[17]{Department of Fire Safety, Seoyeong University, Gwangju 61268, KOREA}
\affil[18]{School of Liberal Arts, Seoul National University of Science and Technology, Seoul, 139-743, KOREA}
\affil[19]{Department of Physics, Soongsil University, Seoul 06978, KOREA}
\affil[20]{Department of Physics, Sungkyunkwan University, Suwon 16419, KOREA}
\affil[21]{Department of Physics and Astronomy, University of Sussex, BN1 9QH, Brighton, UK}
\affil[22]{Research Center for Neutrino Science, Tohoku University, Sendai, Miyagi 980-8577, JAPAN}

\maketitle

\vspace*{-1.5in}
\thispagestyle{empty}

\setlength{\baselineskip}{5mm}
\setlength{\intextsep}{5mm}

\renewcommand{\arraystretch}{0.5}

\newpage

\tableofcontents

\newpage

%%%%%%%%%%%%%%%%%%%%%%%%%%%%%%%%%%%%%%%%%%%%%%%%%%
\section{Introduction}
\indent
%%%%%%%%%%%%%%%%%%%%%%%%%%%%%%%%%%%%%%%%%%%%%%%%%%

This proposal describes the goal and expected sensitivity of the
JSNS$^2$-II at J-PARC Materials and Life
Science Experimental Facility (MLF).

\begin{figure}[h]
 \centering
 \includegraphics[width=0.7 \textwidth]{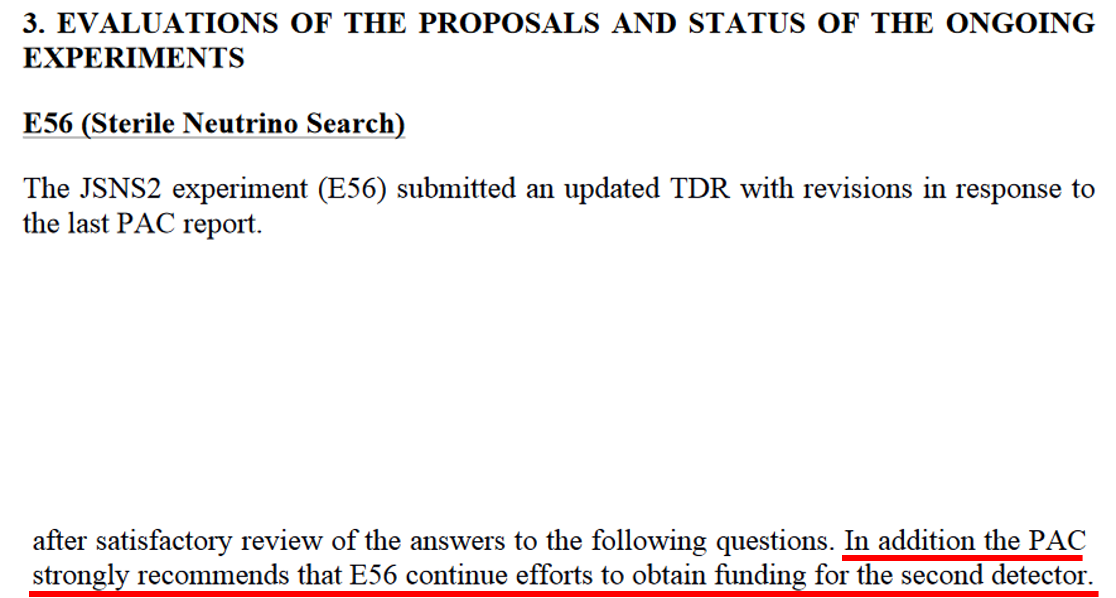}
\caption{\setlength{\baselineskip}{4mm}
  The 25$^{th}$ J-PARC PAC's recommendation. 
}
 \label{Fig:PACRec}
\end{figure}

The JSNS$^2$-II is the second phase of the JSNS$^2$
experiment (J-PARC Sterile Neutrino Search at
J-PARC Spallation Neutron Source)~\cite{CITE:Proposal, CITE:TDR}
with two detectors which are located in 24 m and
48 m baselines to improve the sensitivity of the
search for sterile neutrinos, especially in the low $\Delta m^2$ region, which
has been indicated by the global fit of the appearance mode~\cite{CITE:Review}.
Note that the experiment with two detectors
were proposed in the JSNS$^2$ proposal~\cite{CITE:Proposal}, and 25$^{th}$
J-PARC PAC strongly recommended to make the two detector
configuration~\cite{CITE:25thPAC} (2018) as shown in Fig.~\ref{Fig:PACRec}
even though we started the JSNS$^2$ with one detector during the first phase.
Based on the J-PARC PAC's recommendation, the JSNS$^2$ collaboration has put
a lot of effort in securing the grant, and the Grant-in-Aid for Specially
Promoted Research was awarded in 2020. Therefore, the funding is now available
to build the second detector.

The JSNS$^2$ aims to have a direct test of the LSND experiment~\cite{CITE:LSND}
using the same neutrino source ($\mu$ Decay-At-Rest), neutrino target (proton), 
and detection principle (Inverse-Beta-Decay: IBD) but uses the short-pulsed
3 GeV proton beam and Gd-loaded liquid scintillator (GdLS). Therefore,
100 times better signal-to-noise ratio is given compared to the LSND
experiment.
Please see the reference~\cite{CITE:TDR} for more details such as the
current setup and sensitivity of the JSNS$^2$ experiment.

\begin{figure}[h]
 \centering
 \includegraphics[width=0.7 \textwidth]{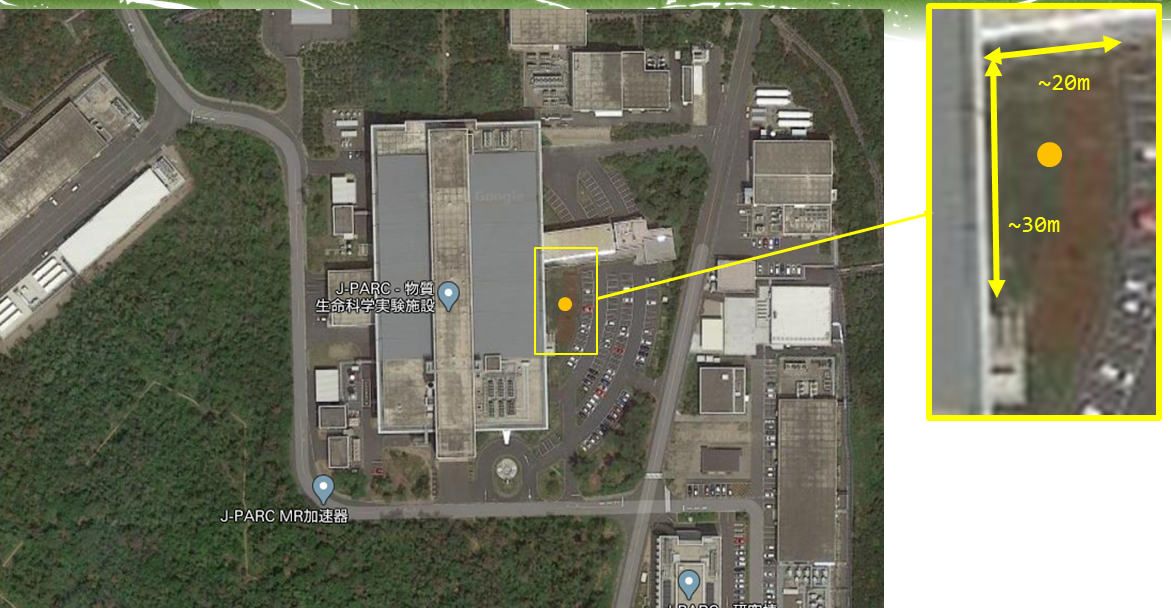}
\caption{\setlength{\baselineskip}{4mm}
  The location of the 2nd detector in the JSNS$^2$-II. The orange circle
  in the picture shows the location of the detector.
}
 \label{Fig:outside}
\end{figure}

The approved Proton-On-Target (POT) of the current JSNS$^2$ with one
detector from the J-PARC PAC is 1.114$\times$10$^{23}$ (1 MW $\times$ 3 years),
while we aim to have 1 MW $\times$ 5 years for the JSNS$^2$-II.
Considering the constraints in the MLF and the Japanese Fire Law,
the baseline of the second detector is 48 m from the target. 
To get a better sensitivity, 
the second detector which has 35 tons of fiducial weight, locates outside
of the MLF as shown in Fig.~\ref{Fig:outside}.
On the other hand, the current existing detector stays at the location
of 24 m baseline even after the second detector construction.

We already started to discuss the issues regarding the Japanese Fire Law and
MLF and there have been no show-stoppers thus far.

%%%%%%%%%%%%%%%%%%%%%%%%%%%%%%%%%%%%%%%%%%%%%%%%%%
\section{Advantage of two detector configuration}
\indent
%%%%%%%%%%%%%%%%%%%%%%%%%%%%%%%%%%%%%%%%%%%%%%%%%%

The detector located in the longer baseline can search for the
neutrino oscillations with the lower $\Delta m^2$ region in general.  
However, in addition, there are a few large advantages to use two detectors
compared to that uses one detector as follows:

\begin{enumerate}
\item If sterile neutrinos exist, the observed 
  oscillation pattern as a function of energy will be changed because
  the neutrino oscillation probability is changed as functions of the
  flight length and energy of neutrinos as: 
  $P(\bar{\nu_{\mu}} \rightarrow \bar{\nu_{e}}) =
  \sin^{2}2\theta \sin^{2}(\frac{1.27 \cdot \Delta m^{2} (eV^{2})
    \cdot L (m)}{E_{\nu} (MeV)})$.
  Figure~\ref{Fig:NOP} shows the difference of the oscillation probabilities
  as a function of the energy with two different baselines with the
  different oscillation parameters.
\begin{figure}[htb]
\centering
\subfigure[24m (LSND best fit)]{
  \includegraphics[width=0.4\textwidth,angle=0]{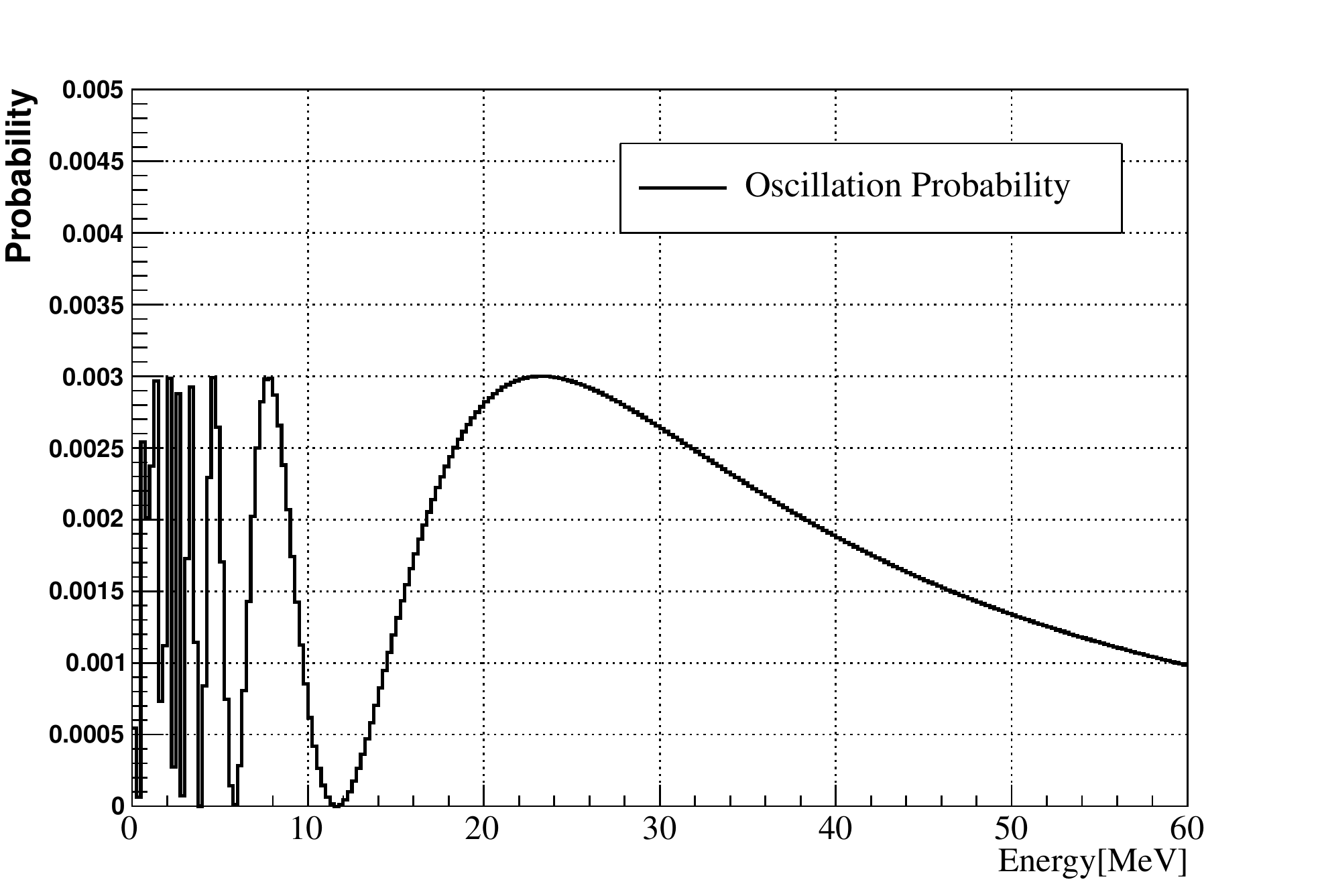}
}
\subfigure[48m (LSND best fit)]{
\includegraphics[width=0.4\textwidth,angle=0]{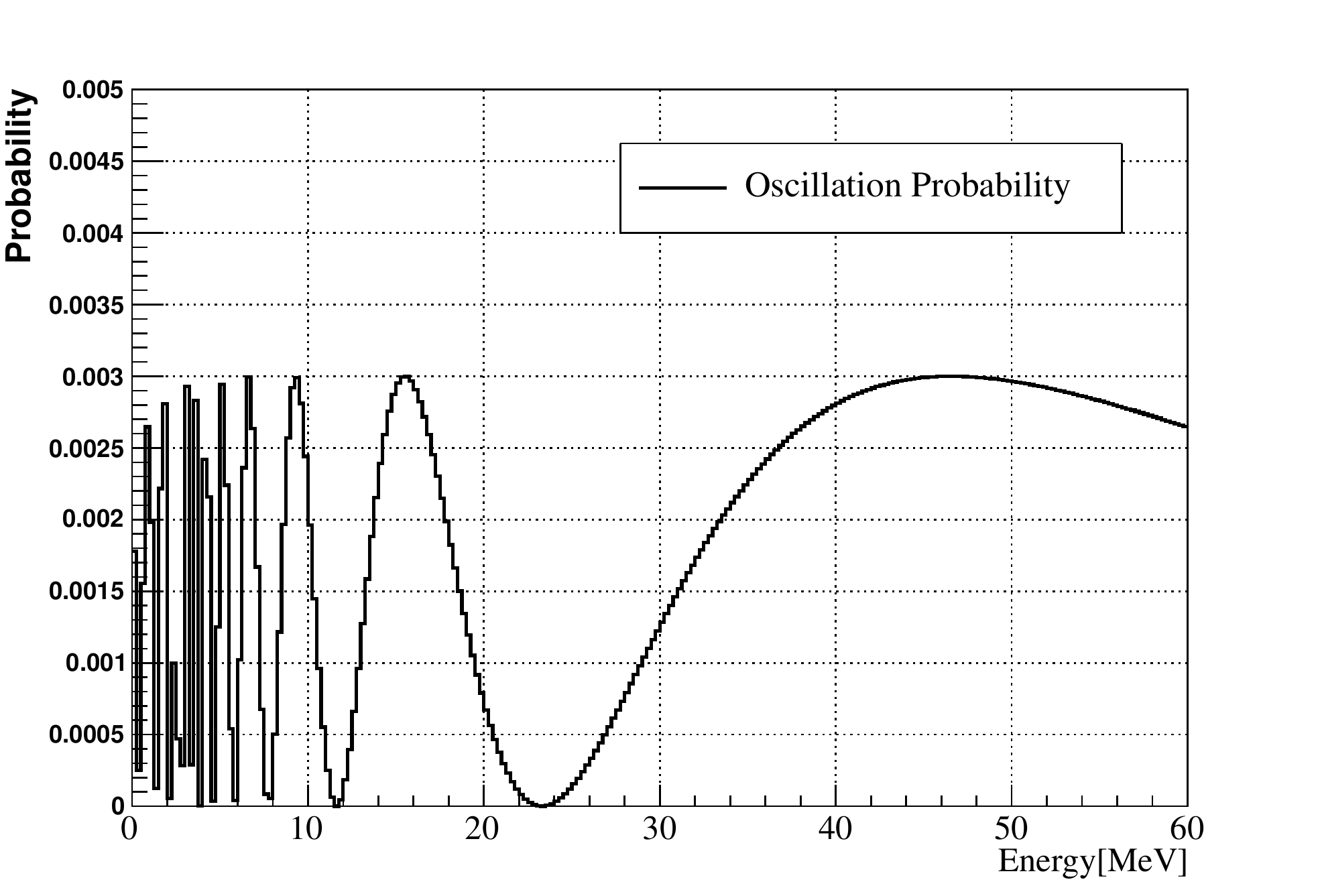}
}
\subfigure[24m (JSNS$^2$ best fit case)]{
\includegraphics[width=0.40\textwidth,angle=0]{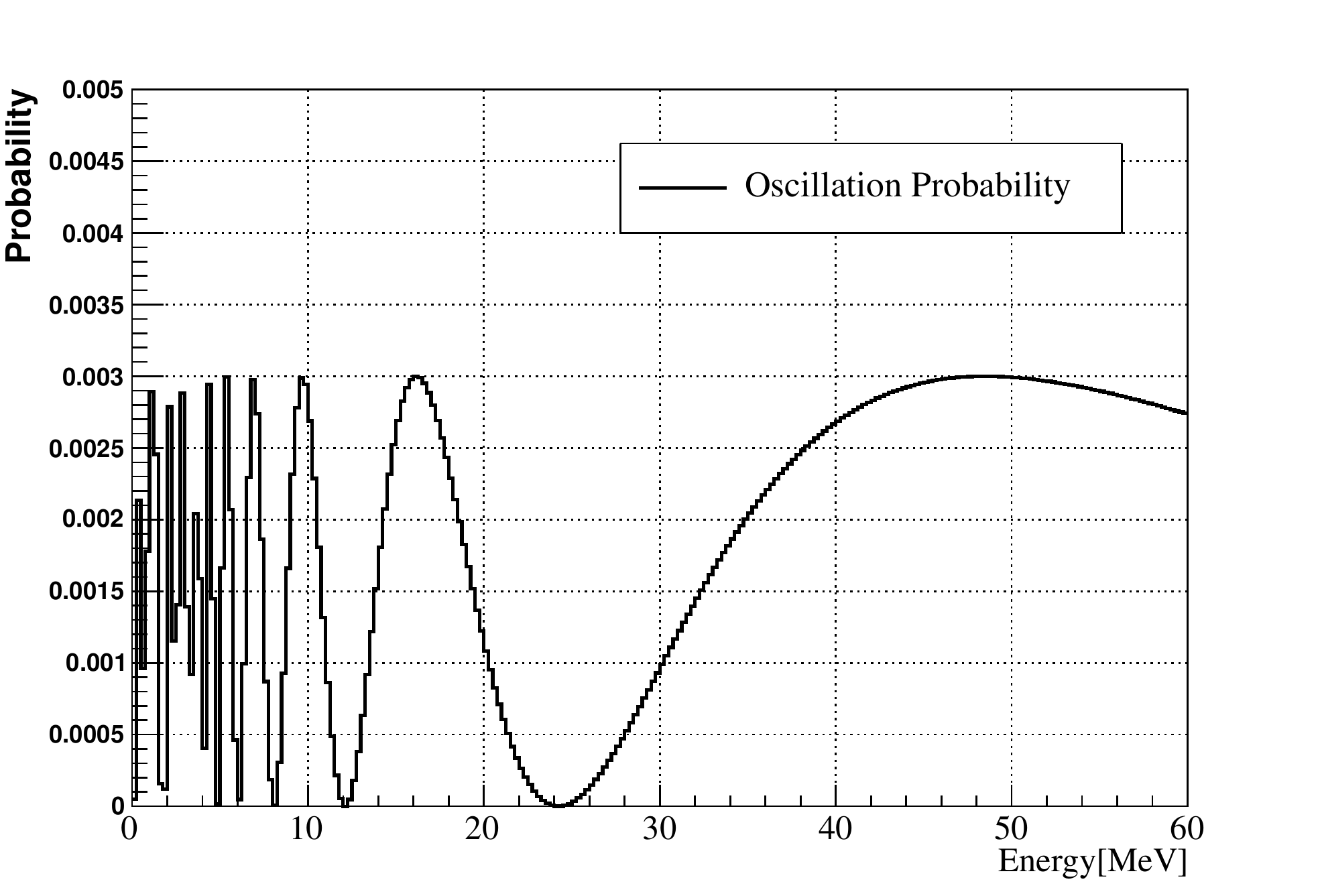}
}
\subfigure[48m (JSNS$^2$ best fit case)]{
\includegraphics[width=0.40\textwidth,angle=0]{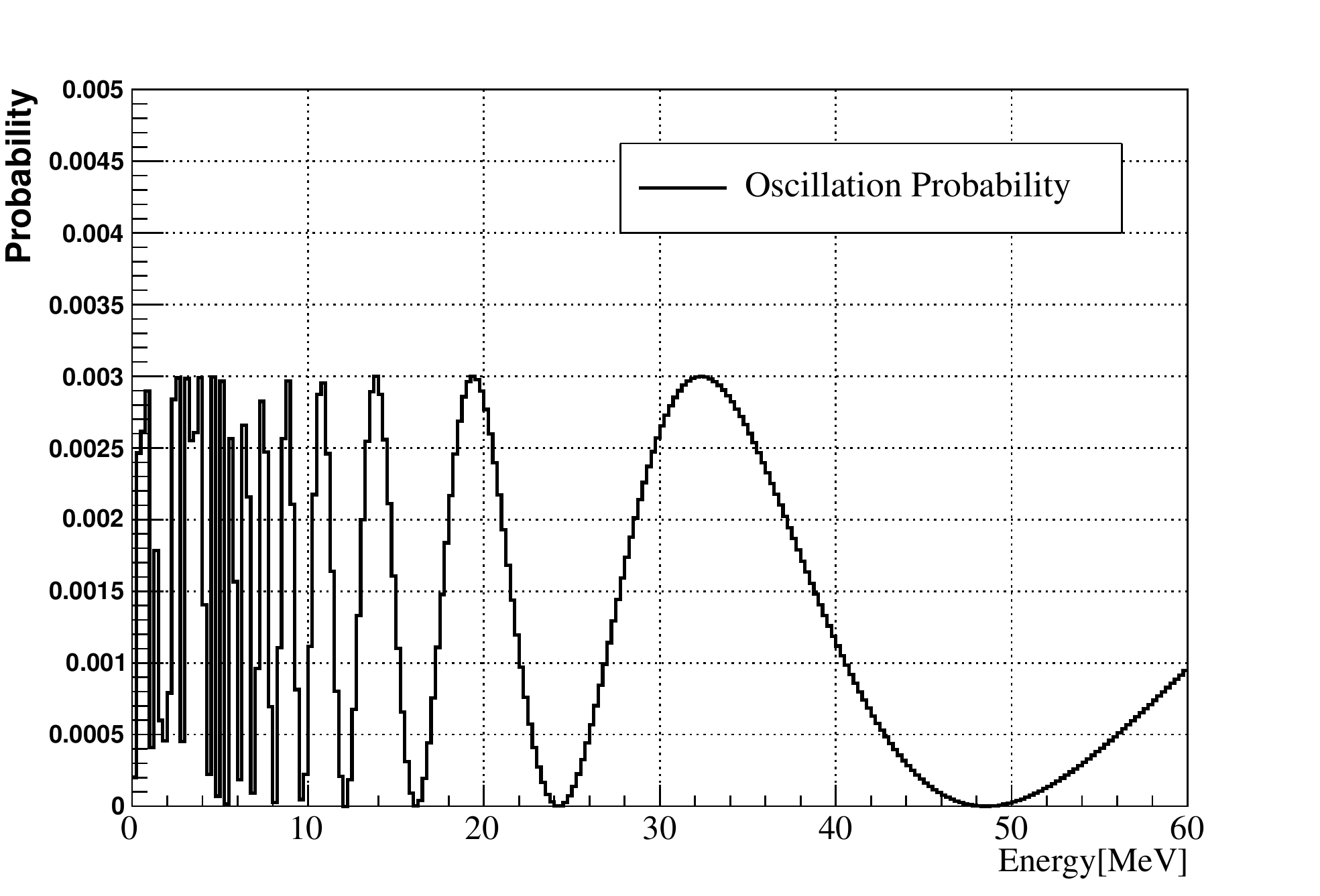}
}
\caption{\setlength{\baselineskip}{4mm}
  Oscillation probabilities at 24 m (left) and 48 m (right) with
  two different oscillation paramaters. Top: LSND best fit parameters
  ($\Delta m^2, sin^2 (2\theta)$)=(1.2, 0.003), bottom: the JSNS$^2$
  single detector best fit case. ($\Delta m^2, sin^2 (2\theta)$)=(2.5, 0.003)
} 
\label{Fig:NOP}
\end{figure}
\item The background rates are different between two detectors because
  the backgrounds induced from the beam are reduced as a function
  of distance (1/r$^2$: r is the distance from the mercury target), while
  backgrounds induced by cosmic rays does not depend on the distance.
  This makes much better understandings of the backgrounds compared to
  the one-detector scheme.
\item Systematic uncertainties such as normalization of neutrino fluxes
  are canceled out using two detectors. This effect will be shown
  later (section~\ref{Sec:ES}).
\end{enumerate}
  
It might be ideal to use two identical detectors. However
JSNS$^2$-II aims to look for the $\bar{\nu}_{\mu} \to \bar{\nu}_{e}$ appearance
and the requirements to the systematic uncertainties are
relatively small. Therefore, after the detailed understanding of
the detector, it is possible to achieve our goals even using
the different size detectors.

%%%%%%%%%%%%%%%%%%%%%%%%%%%%%%%%%%%%%%%%%%%%%%%%%%
\section{Recent status of sterile neutrino searches}
\indent
%%%%%%%%%%%%%%%%%%%%%%%%%%%%%%%%%%%%%%%%%%%%%%%%%%

The situation up to 2015 was described in the JSNS$^2$ Technical
Design Report (TDR)~\cite{CITE:TDR}.
There have been several new results since then.
Here we explain the relevant ones, but please see
the reference which reviews them in details~\cite{CITE:Review}.

%%%%%%%%%%%%%%%%%%%%%%%%%%%%%%%%%%%%%%%%%%%%%%%%%%
\subsection{World-wide status}
\indent
%%%%%%%%%%%%%%%%%%%%%%%%%%%%%%%%%%%%%%%%%%%%%%%%%%

%%%%%%%%%%%%%%%%%%%%%%%%%%%%%%%%%%%%%%%%%%%%%%%%%%
\subsubsection{$\nu_{\mu} \to \nu_{e}$ appearance mode}
\indent
%%%%%%%%%%%%%%%%%%%%%%%%%%%%%%%%%%%%%%%%%%%%%%%%%%

The MiniBooNE experiment updated their results recently:
Their latest publication~\cite{CITE:MB} mentioned that the significance of
the $\bar{\nu_{\mu}} \to \bar{\nu_{e}}$ signal compared to background
only is 4.7$\sigma$ now.
However, as the MicroBooNE experiment~\cite{CITE:MicroB} pointed out, 
the observed excess of events in the low-energy region could be
due to single $\gamma$ production interactions, which are poorly
understood in the current theory, therefore the MiniBooNE cannot
distinguish between the oscillation signal and this background because
it is a Cherenkov detector. The MicroBooNE uses a liquid argon
TPC detector and they can distinguish the oscillation signal
and single $\gamma$ events. They are expected to publish
their results in near future.

The far detector of the Short Baseline Neutrino experiment (SBN) at FNAL,
the ICARUS detector, is ready to take data.
They are already fully filled with liquid argon~\cite{CITE:ICARUS},
and are waiting for beam. The SBN experiment will provide direct
confirmation or refutal of MiniBooNE anormaly.

%%%%%%%%%%%%%%%%%%%%%%%%%%%%%%%%%%%%%%%%%%%%%%%%%%
\subsubsection{Disappearance mode}
\indent
%%%%%%%%%%%%%%%%%%%%%%%%%%%%%%%%%%%%%%%%%%%%%%%%%%

If the neutrino oscillation due to fourth or more mass eigenstates
(i.e. sterile neutrino(s)) exist,
the $\nu_{e} \to \nu_{s}$ and $\nu_{\mu} \to \nu_{s}$ are also happened
accordingly. Especially, three active plus one sterile neutrino model have a
simple relationship between appearance and disappearance modes, as
shown in our TDR~\cite{CITE:TDR} and many other publications
(see for example \cite{CITE:Review}).

For the $\nu_{e} \to \nu_{s}$ searches using the $\nu_{e}$ disappearance,
there have been (super-)short
baseline reactor experiments, such as Daya-Bay~\cite{CITE:DB},
RENO~\cite{CITE:RENO}, DANSS~\cite{CITE:DANSS}, NEOS~\cite{CITE:NEOS},
Neutrino-4~\cite{CITE:N4},
PROSPECTS~\cite{CITE:PROSPECTS}, Stereo~\cite{CITE:STEREO}.
They typically put the scintillator
detectors with the baseline of $\sim$10 meters. 
The Neutrino-4 experiment declares that they observed a clear oscillation
pattern, while
other experiments showed the exclusion regions for the two-dimensional
mixing angles and $\Delta m^2$ plane
although some of allowed regions are remained. 

$\nu_{\mu} \to \nu_{s}$ searches using the $\nu_{\mu}$ disappearance
have been developed by IceCube~\cite{CITE:IC},
Super-Kamionkande~\cite{CITE:SK}, MINOS/MINOS+~\cite{CITE:MINOS} and
T2K~\cite{CITE:T2K} recently.
Their results showed no sterile neutrinos, and there is clear tension
between the apperance and the disapperance modes.

%%%%%%%%%%%%%%%%%%%%%%%%%%%%%%%%%%%%%%%%%%%%%%%%%%
\subsection{The current JSNS$^2$ experiment}
\indent
%%%%%%%%%%%%%%%%%%%%%%%%%%%%%%%%%%%%%%%%%%%%%%%%%%

Under the current situation, the experiments that provide direct tests
for the apperance modes are getting crucial. For the direct test of
MiniBooNE, MicroBooNE itself and SBN are important.
On the other hand, for the direct test of LSND, the JSNS$^2$
plays an important role. The second detector of the JSNS$^2$ must be constructed
in a timely fashion because the funding to build the second detector was
granted in 2020 as recommended by J-PARC PAC.

The JSNS$^2$ has started data taking from June 2020~\cite{CITE:JSNS2}.
Currently, we are checking the background rates in the data.
We will show some concrete background numbers in the presentation
at the next J-PARC PAC.

Currently, we assume that the amount of backgrounds in the second detector
is the same as the current one although we expect that they are much
smaller backgrounds
because the beam-related background, such as gamma rays from
MLF third floor concrete, will be reduced drastically compared to
the current JSNS$^2$~\cite{CITE:TDR}.

%%%%%%%%%%%%%%%%%%%%%%%%%%%%%%%%%%%%%%%%%%%%%%%%%%
\section{The second detector}
\indent
%%%%%%%%%%%%%%%%%%%%%%%%%%%%%%%%%%%%%%%%%%%%%%%%%%

The location of the second detector is shown in Fig.~\ref{Fig:outside}.
In this location, we can have a 8 m $\times$ 8 m space to satisfy
the Japanese Fire Law constraint, which requires an empty space of 5 m
from the edge of the detector.

\begin{figure}[h]
 \centering
 \includegraphics[width=0.7 \textwidth]{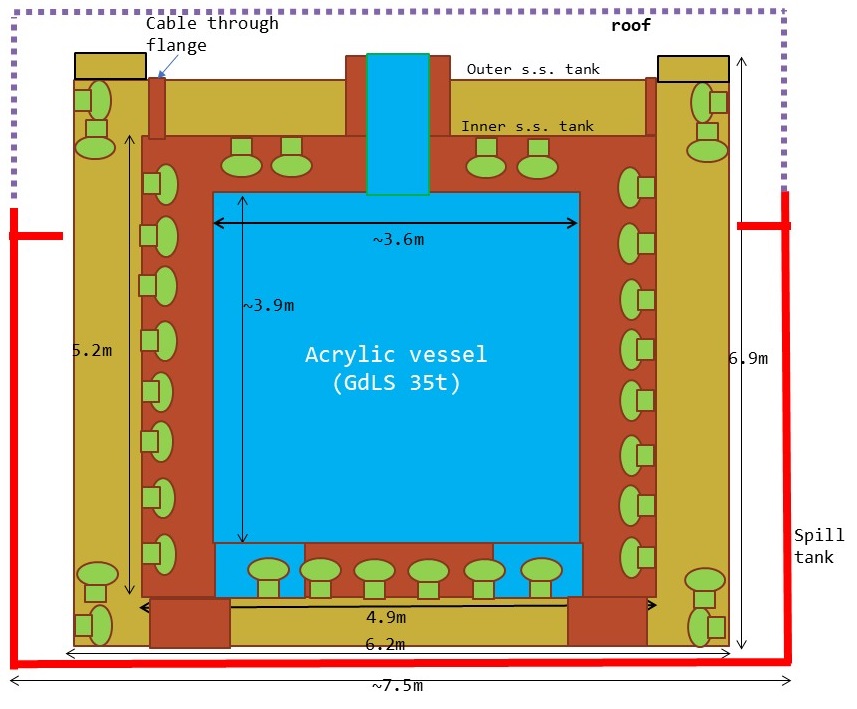}
\caption{\setlength{\baselineskip}{4mm}
  The structure of the 2nd detector in the JSNS$^2$-II. 
}
 \label{Fig:2Det}
\end{figure}

The second detector has a similar structure as the existing JSNS$^2$
detector, which is already working.
The conceptual design of the detector is shown in Fig.~\ref{Fig:2Det}.
As shown here, the detector is placed outside the MLF building and
no specific buildings for the detector is made, therefore,
we will make 
a stainless roof on the top of the detector. The electronics, PMT cables and
slow monitor/control system are put here.

To compensate for the reduction of the neutrino flux due to the distance
from the mercury target, the target mass of the GdLS,
which is the Linear AlkylBenzene (LAB) based liquid scintillator, 
inside the acrylic vessel is increased to 35 tons.
On the other hand, the detector located at the longer distance is better at
searching in the lower $\Delta m^2$ region of the neutrino oscillation.

Similar to reactor experiments such as Daya-Bay~\cite{CITE:DB},
RENO~\cite{CITE:RENO2}, Double-Chooz~\cite{CITE:DC},
we will have double stainless tank structure. The inner stainless
structure makes the sepration of gamma catcher (dark brown) and veto
(light brown) regions.
In veto region, the same reflector sheets (REIKO
LUIREMIRROR~\cite{CITE:REIKO}) as the existing JSNS$^2$ detector
will be used.  
The gamma catcher and veto regions are filled with
pure liquid scintillator (LS),
which is also based on the LAB. The total weight of the pure LS
is about 130 tons.

To keep the same photo coverage of the detector as the first detector,
we will surround the acrlyic vessel with the 240 PMTs. 

All details of this detector structure will be described in the Technical
Design Report (TDR) of the JSNS$^2$-II. Some of GdLS and pure LS (and PMTs
if possible) could be
donated from the Daya-Bay experiment. The JSNS$^2$ has started and
will continue to negotiate for this donation.

%%%%%%%%%%%%%%%%%%%%%%%%%%%%%%%%%%%%%%%%%%%%%%%%%%
\section{Signal and backgrounds}
\indent
%%%%%%%%%%%%%%%%%%%%%%%%%%%%%%%%%%%%%%%%%%%%%%%%%%

Based on the JSNS$^2$ TDR~\cite{CITE:TDR}, only the following three data samples
must be considered to calculate the sensitivity. Other background components
are negligible. The realistic estimation of the background
components with the real data taken by the existing detector
will be shown in the PAC presentation. 

\begin{enumerate}
\item signal : $\mu^{+} \to e^{+} + \nu_{e} + \bar{\nu}_{\mu} : \bar{\nu}_{\mu} \to \bar{\nu}_{e}$. 
\item intrinsic background : $\mu^{-} \to e^{-} + \nu_{\mu} + \bar{\nu}_{e}$.
\item accidental backgrounds 
\end{enumerate}

Hereafter, we will assume that the numbers of events with the event
selection efficiency
are the same as
those in the TDR if the time exposure, detector size, and the baseline of the
location are identical,
but scaled as function of exposure time and detector fiducial
volume for the accidental background, and scaled by 1/r$^2$
in addition to the time exposure and
detector size, for the intrinsic background.

Table~\ref{TAB:grandsum} summarizes the number of events in the
TDR and those in this proposal configuration.
The 1 MW $\times$ 5 years are assumed here to conclude LSND region
with 3 sigma confidence level.

\begin{table}[ht]
\begin{center}
\begin{tabular}{|c|c|c|c|c|}\hline
& Contents & Current JSNS$^2$ & near(24m) & far(48m)\\
& & 17tons & 17tons & 35tons\\
& & 5000h/y$\times$3y & 5000h/y$\times$5y & 5000h/y$\times$5y \\ \hline
&$sin^22\theta=3.0\times10^{-3}$& & & \\
&$\Delta m^2=2.5eV^{2}$ & 87 & 145 & 24\\ 
&&&& \\ \cline{2-5} 
&$sin^22\theta=3.0\times10^{-3}$&&&\\
&$\Delta m^2=1.2eV^{2}$& 62 & 103 & 77\\
&(Best fit values of LSND)&&&\\\hline\hline
&$\overline{\nu}_{e}$ from $\mu^{-}$ & 43 & 72 & 36 \\\cline{2-5}
&$^{12}C(\nu_{e},e^{-})^{12}N_{g.s.}$& 3 & 5 & 2 \\\cline{2-5}
background&beam-associated fast n & $\le$2 & $\le$3 & $\le$2\\\cline{2-5}
&Cosmic-induced fast n& negligible & negligible & negligible \\\cline{2-5}
&Total accidental events & 20 & 33 & 67 \\\hline
\end{tabular}
\caption{\setlength{\baselineskip}{4mm}
  Summary of the number of events based on the reference~\cite{CITE:TDR}.
  Summary of the event rate for
  5000 h/y$\times$3 years for one 17-ton detector (left:~\cite{CITE:TDR})
  and those with 5000 h/y$\times$5 years for the near (24 m) and far (48 m)
  detectors. Note that the rate of the beam-related background components
  has been changed as function of the fiducial mass, the exposure time and
  the distance from the target. Cosmic ray related backgrounds are
  changed as function of the fiducial mass and exposure time only.
}
\label{TAB:grandsum}
\end{center}
\end{table}

The expected energy spectra of the background will be described in the
next section.

%%%%%%%%%%%%%%%%%%%%%%%%%%%%%%%%%%%%%%%%%%%%%%%%%%
\section{Energy spectra of signal and backgrounds}
\indent
%%%%%%%%%%%%%%%%%%%%%%%%%%%%%%%%%%%%%%%%%%%%%%%%%%

As described in the JSNS$^2$ proposal~\cite{CITE:Proposal},
the energy spectra of the signal and background are used
to separate them. The beam timing can select the neutrino
production from muon (plus) decay-at-rest ($\mu$DAR) from others.
Note that the Michel spectrum of $\mu$DAR 
is well known thanks to the huge efforts made by the elementary particle
physicists' so far~\cite{CITE:PDG}.

To make the fitting templates, we used the official JSNS$^2$ simulation (RAT)
and reconstruction (JADE) tools.
For example, the energy loss in the acrylic vessel is very well simulated using
GEANT4~\cite{CITE:G4}.

\begin{figure}[htbp]
\centering
\subfigure[The energy spectrum of signal]{
\includegraphics[width=0.45\textwidth,angle=0]{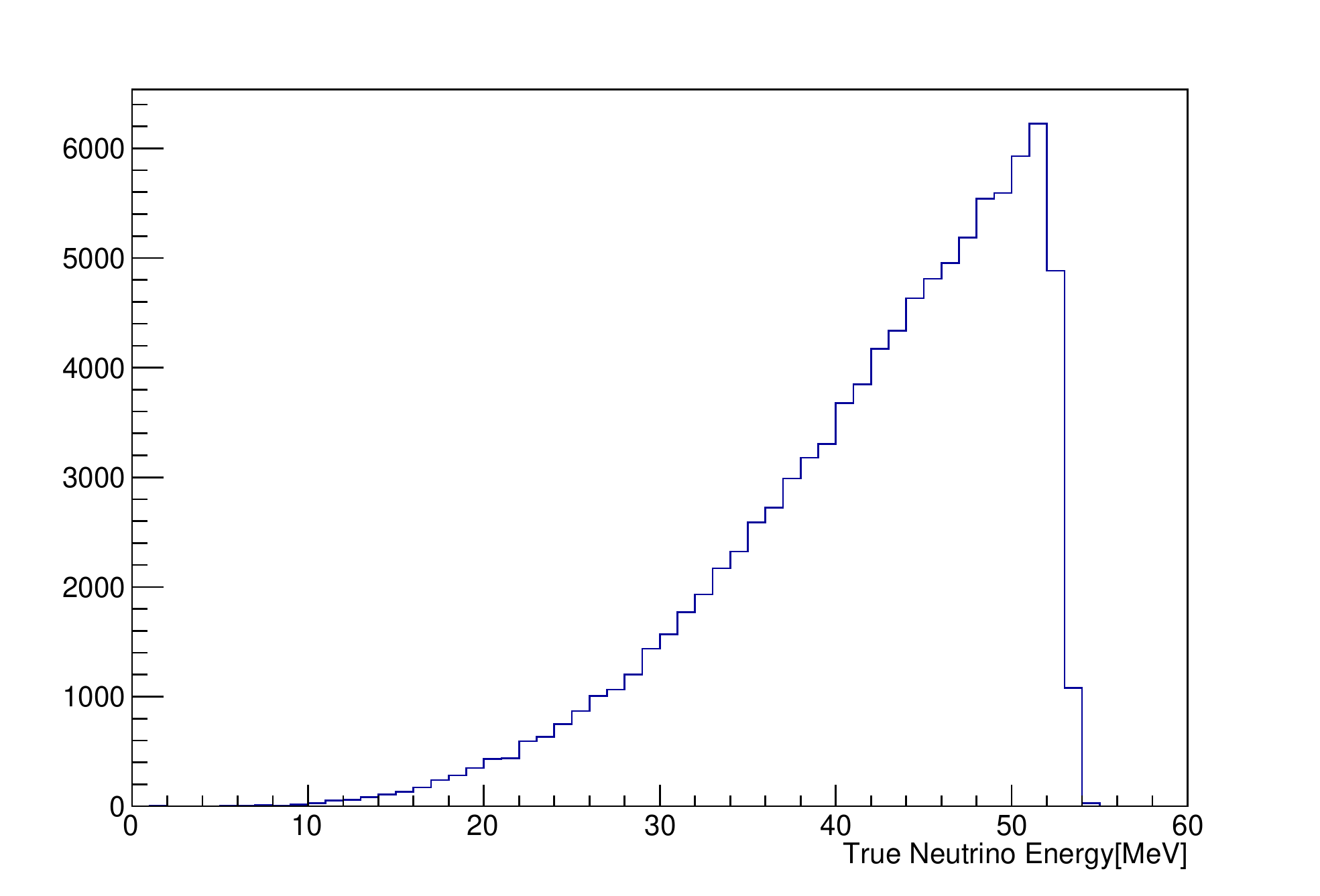}
}
\subfigure[The reconstructed energy of signal events]{
\includegraphics[width=0.45\textwidth,angle=0]{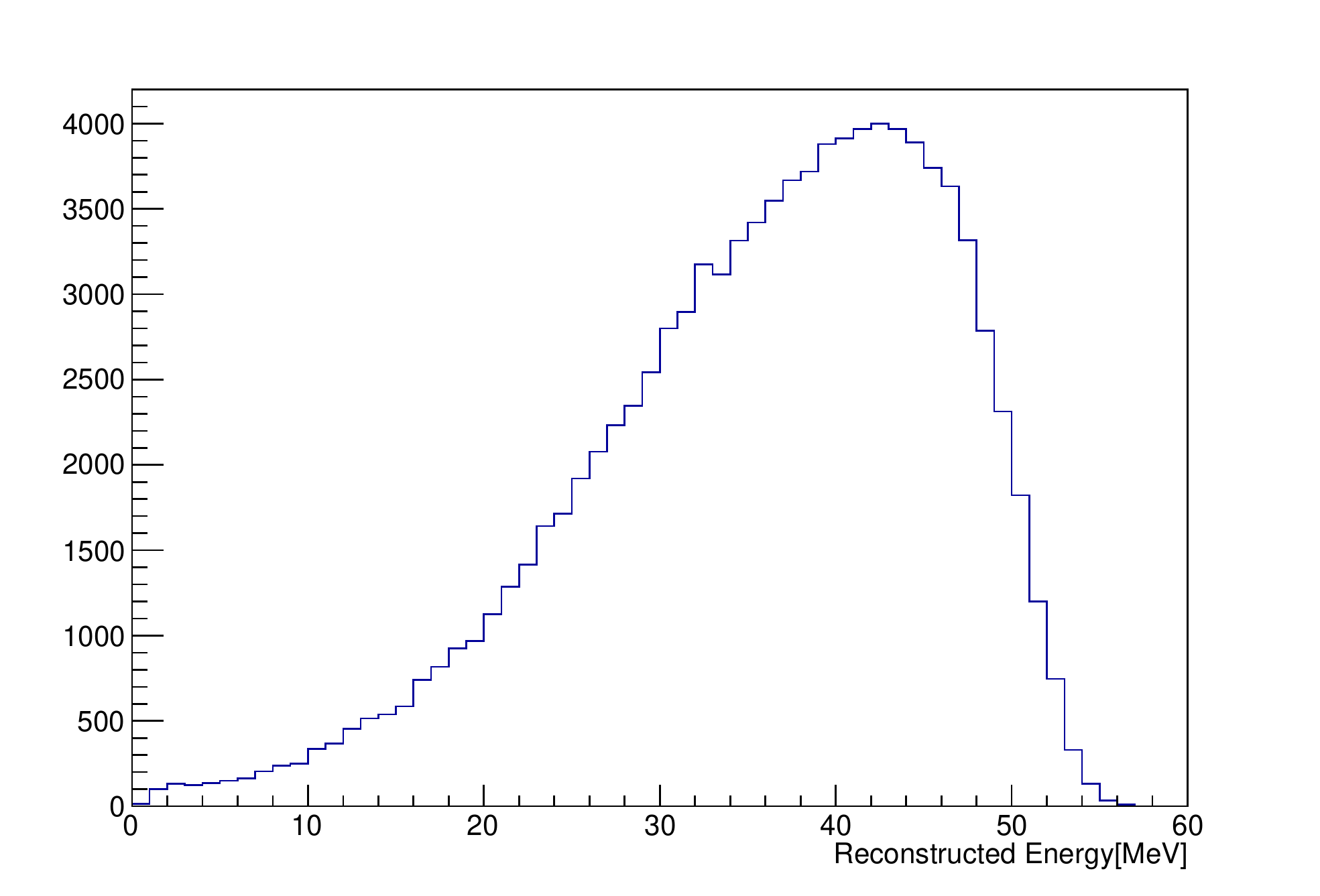}
}
\caption{\setlength{\baselineskip}{4mm}
  True $\bar{\nu}_{e}$ energy (left) and the reconstructed energy (right). 
} 
\label{FIG:signal}
\end{figure}

Figure~\ref{FIG:signal} shows the true $\bar{\nu_{e}}$ energy and reconstructed
energy spectra for the signal (with high $\Delta m^{2}$ region). 
The event vertices are uniformly randomized in the acrylic vessel region of the
detector.
The difference of the spectra between the true $\bar{\nu_{e}}$ energy
and observed reconstructed energy is mainly due to the
energy loss in the acrylic vessel. However, some of them are due to
IBD interaction~\cite{CITE:IBD}. 

\begin{figure}[htbp]
\centering
\subfigure[True $\bar{\nu}_{e}$ vs True Positron Energy]{
\includegraphics[width=0.42\textwidth,angle=0]{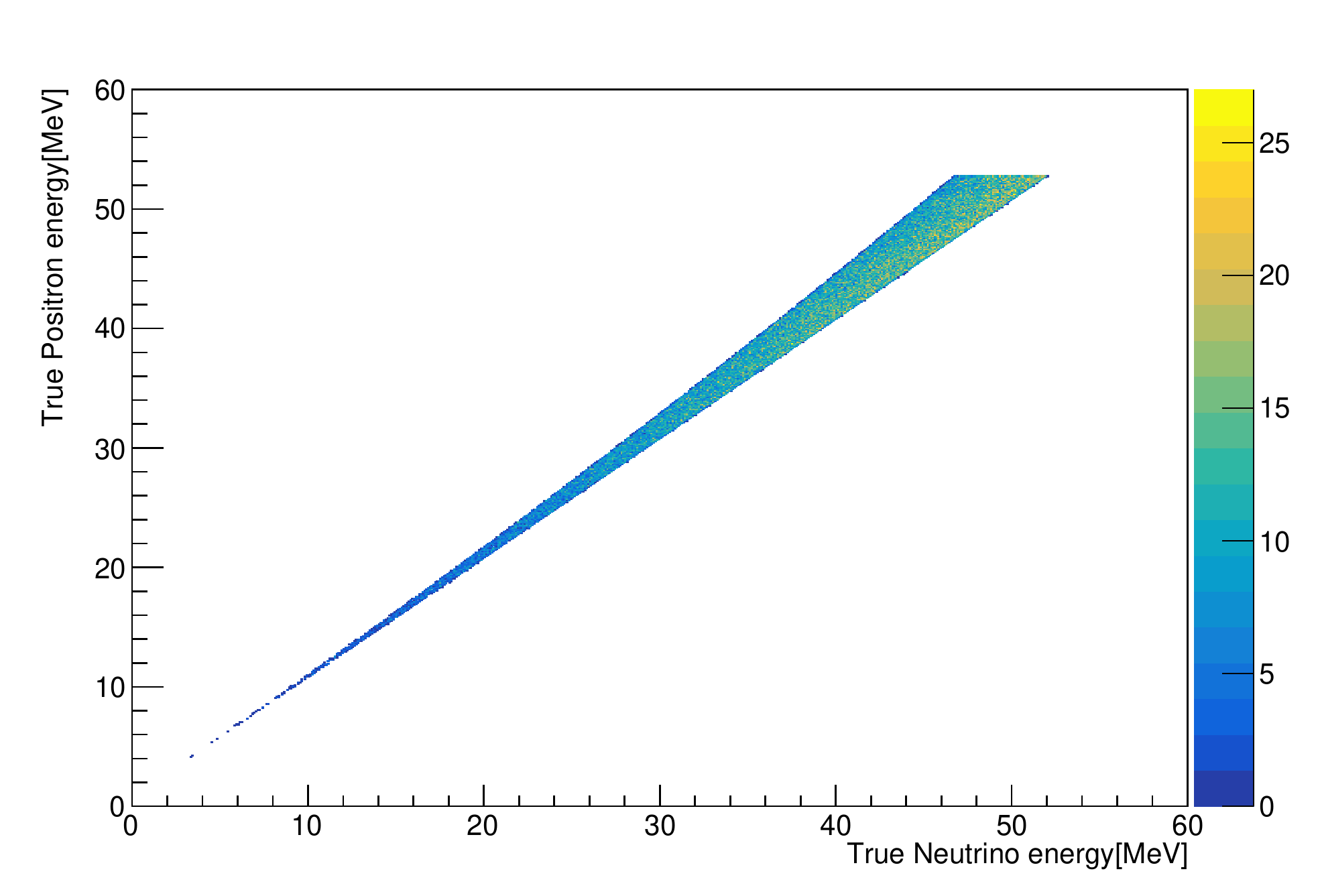}
}
\subfigure[True $\bar{\nu}_{e}$ vs Reconstructed Positron Energy]{
\includegraphics[width=0.45\textwidth,angle=0]{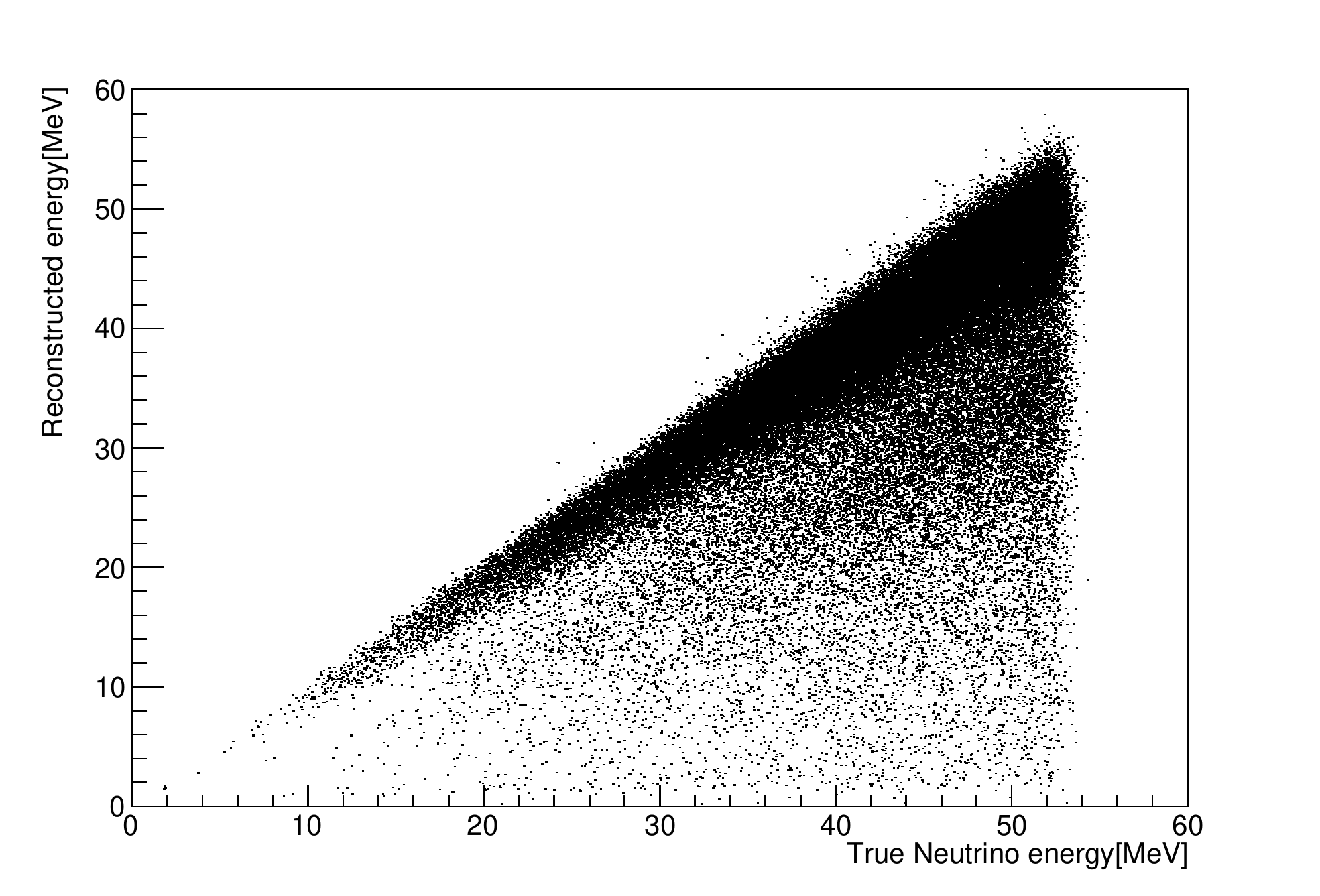}
}
\caption{\setlength{\baselineskip}{4mm}
True energy of $\bar{\nu}_{e}$ (horizontal axis) and true energy of positrons (vertical) simulated one (left). True energy of $\bar{\nu}_{e}$ (horizontal axis) and reconstructed energy of positrons (vertical) (right). 
} 
\label{Fig:EnuvsErec}
\end{figure}

Figure~\ref{Fig:EnuvsErec} shows the 2D plot
between the true positron energy and the true $\bar{\nu_{e}}$ energy (left),
and that between the reconstructed energy and the true $\bar{\nu_{e}}$ energy
(right).
You can see the effects from the IBD neutrino interaction, the energy
reconstruction and the energy loss inside the acrylic vessel.

\begin{figure}[htbp]
\centering
\subfigure[\setlength{\baselineskip}{4mm} The energy spectrum of intrinsic $\bar{\nu}_{e}$ background]{
\includegraphics[width=0.45\textwidth,angle=0]{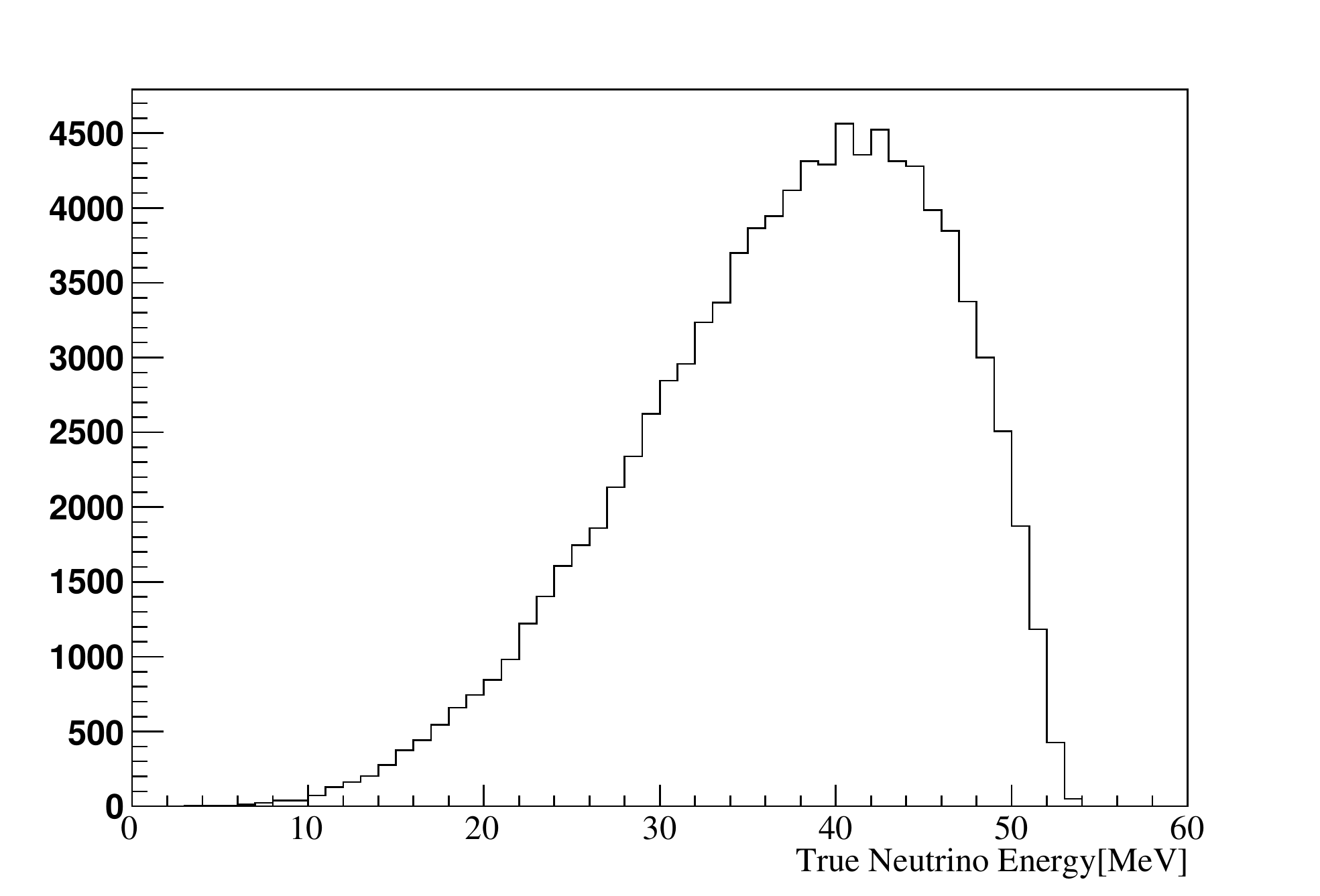}
}
\subfigure[The reconstructed energy]{
\includegraphics[width=0.45\textwidth,angle=0]{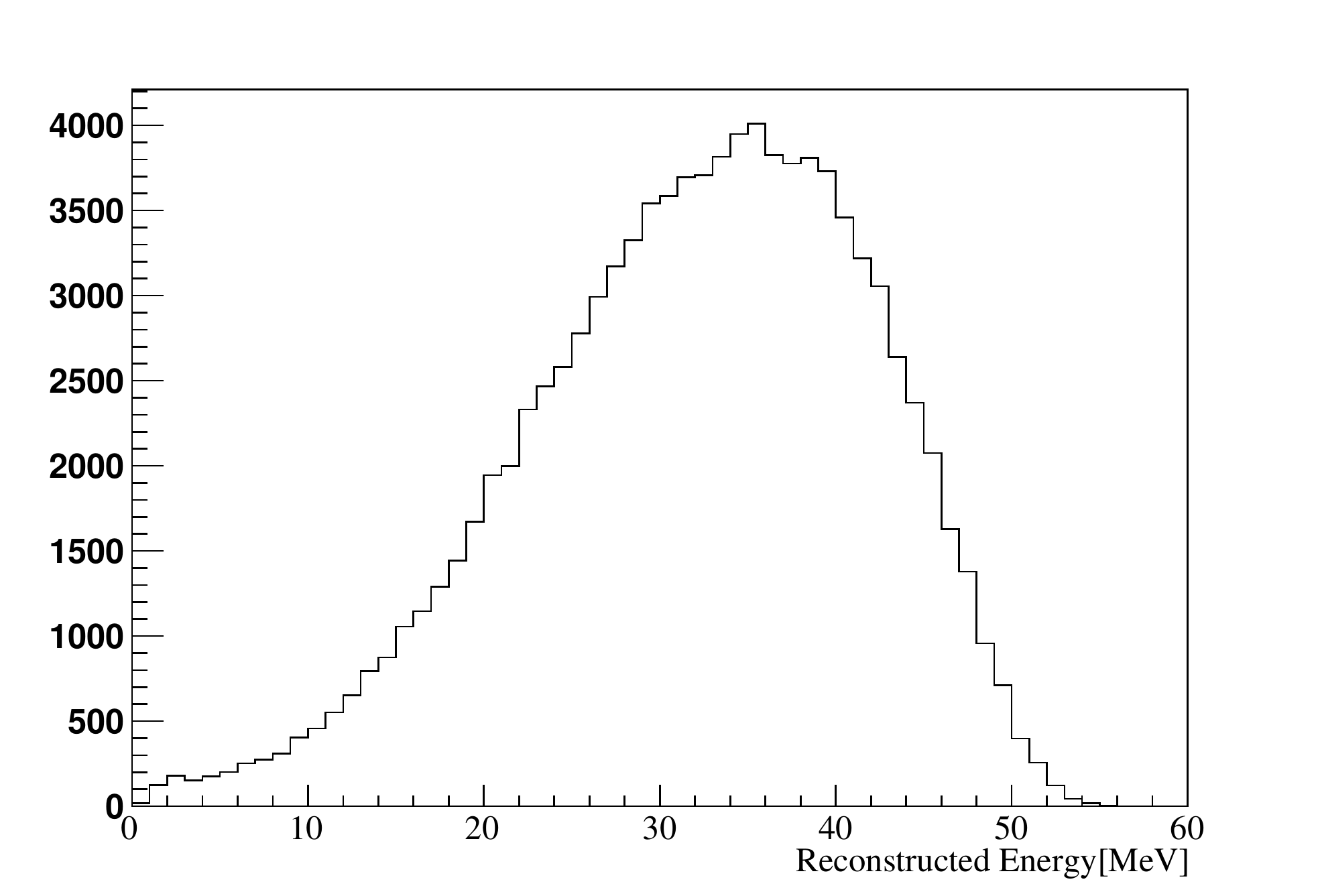}
}
\caption{\setlength{\baselineskip}{4mm}
  True $\bar{\nu}_{e}$ energy (left) and the reconstructed energy (right). 
} 
\label{FIG:intrinsicBKG}
\end{figure}

Figure~\ref{FIG:intrinsicBKG} shows the true and reconstructed
energy spectra for the intrinsic background ($\bar{\nu}_{e}$ from $\mu^{-}$). 
Event vertices are uniformly randomized in the acrylic vessel region.

\begin{figure}[h]
 \centering
 \includegraphics[width=0.6 \textwidth]{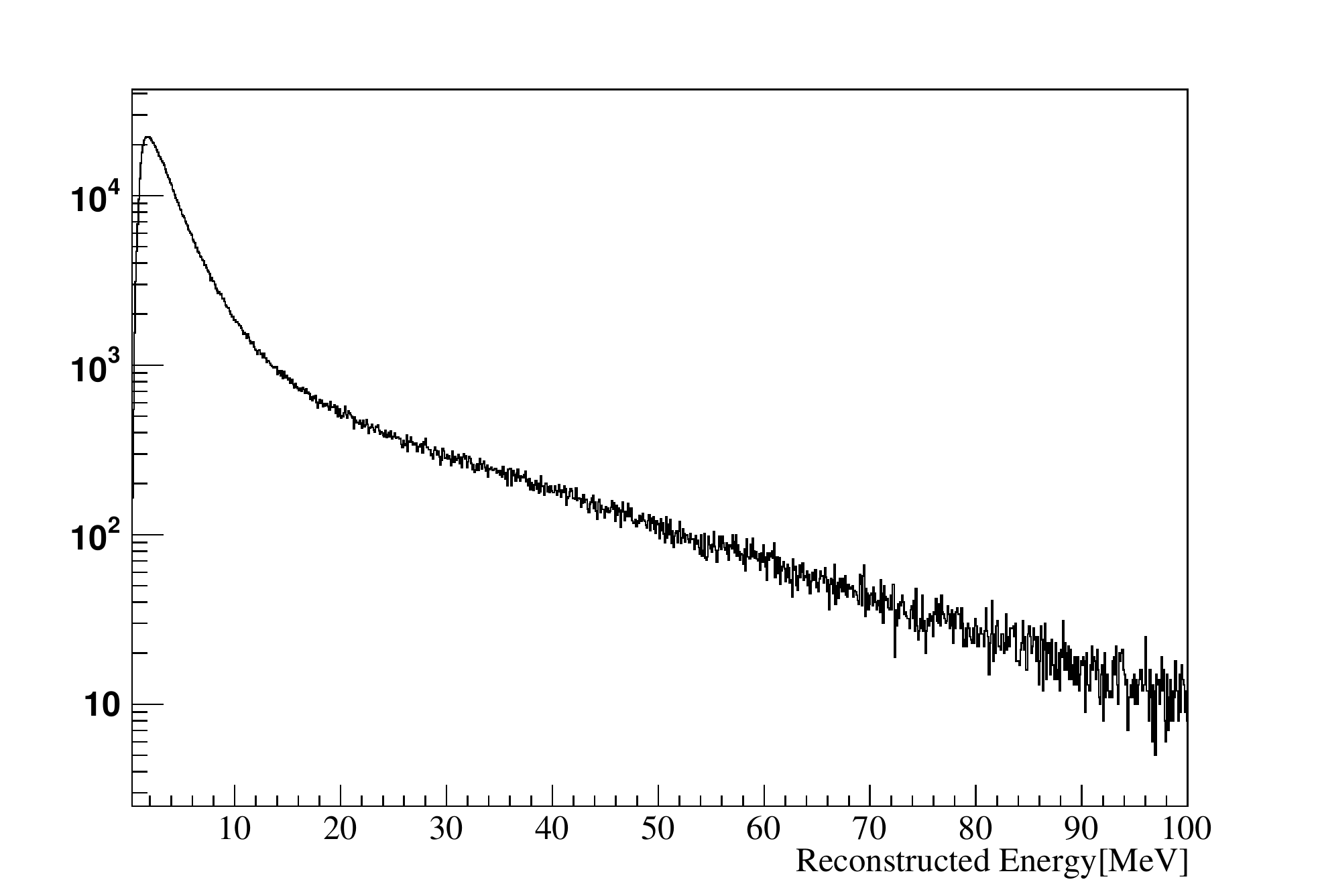}
\caption{\setlength{\baselineskip}{4mm}
The reconstructed energy of the accidental background. 
}
 \label{Fig:CG}
\end{figure}

Figure~\ref{Fig:CG} shows the reconstructed
energy spectrum for the accidental background. The background of
the prompt IBD region is dominated by the gamma rays induced by
cosmic rays.
As pointed out by \cite{CITE:SR_14NOV1}, the gamma rays are generated
on the surface of the stainless steel tank.

\begin{figure}[htbp]
\centering
\subfigure[The energy spectrum of 24m detector]{
\includegraphics[width=0.45\textwidth,angle=0]{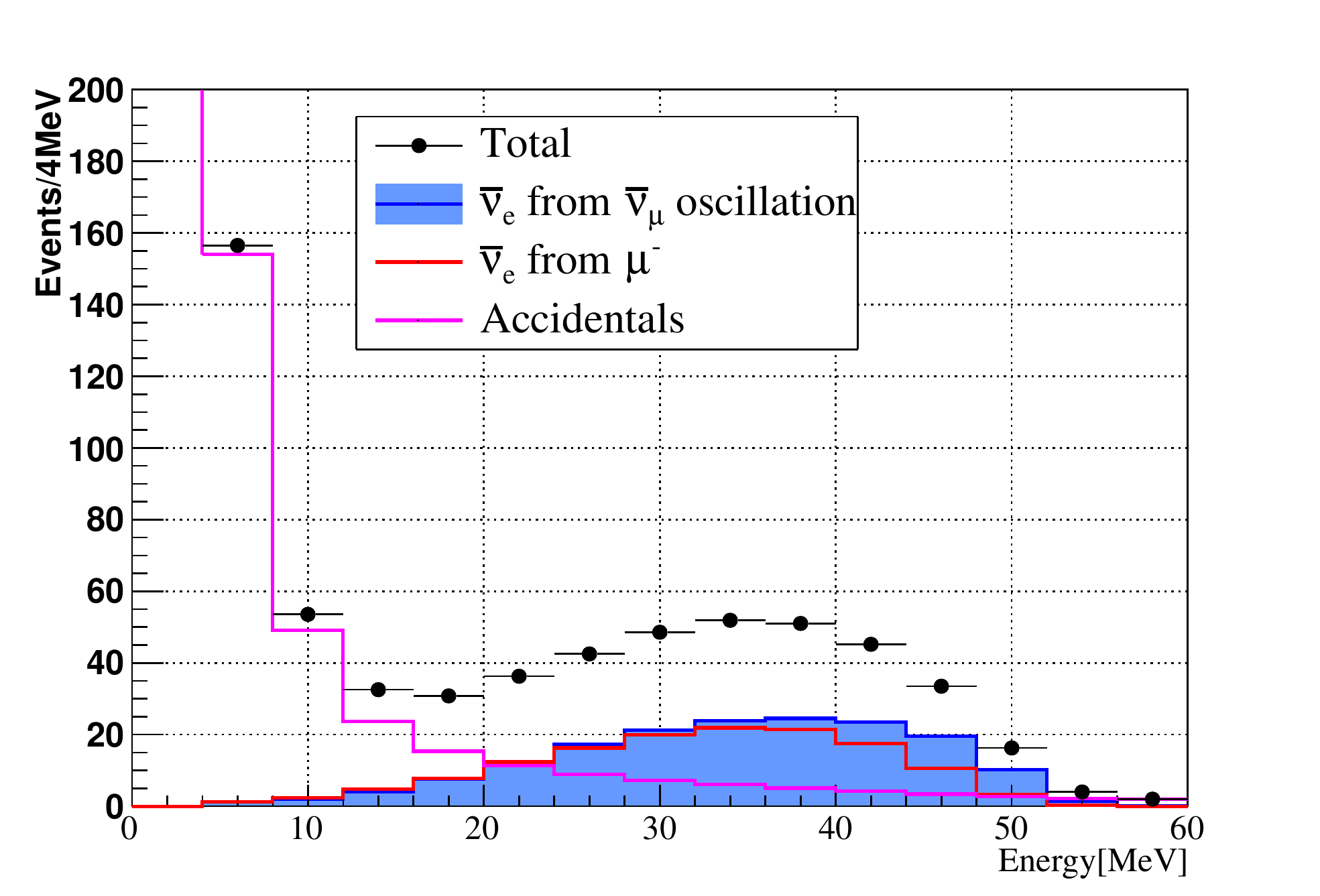}
}
\subfigure[The energy spectrum of 48m detector]{
\includegraphics[width=0.45\textwidth,angle=0]{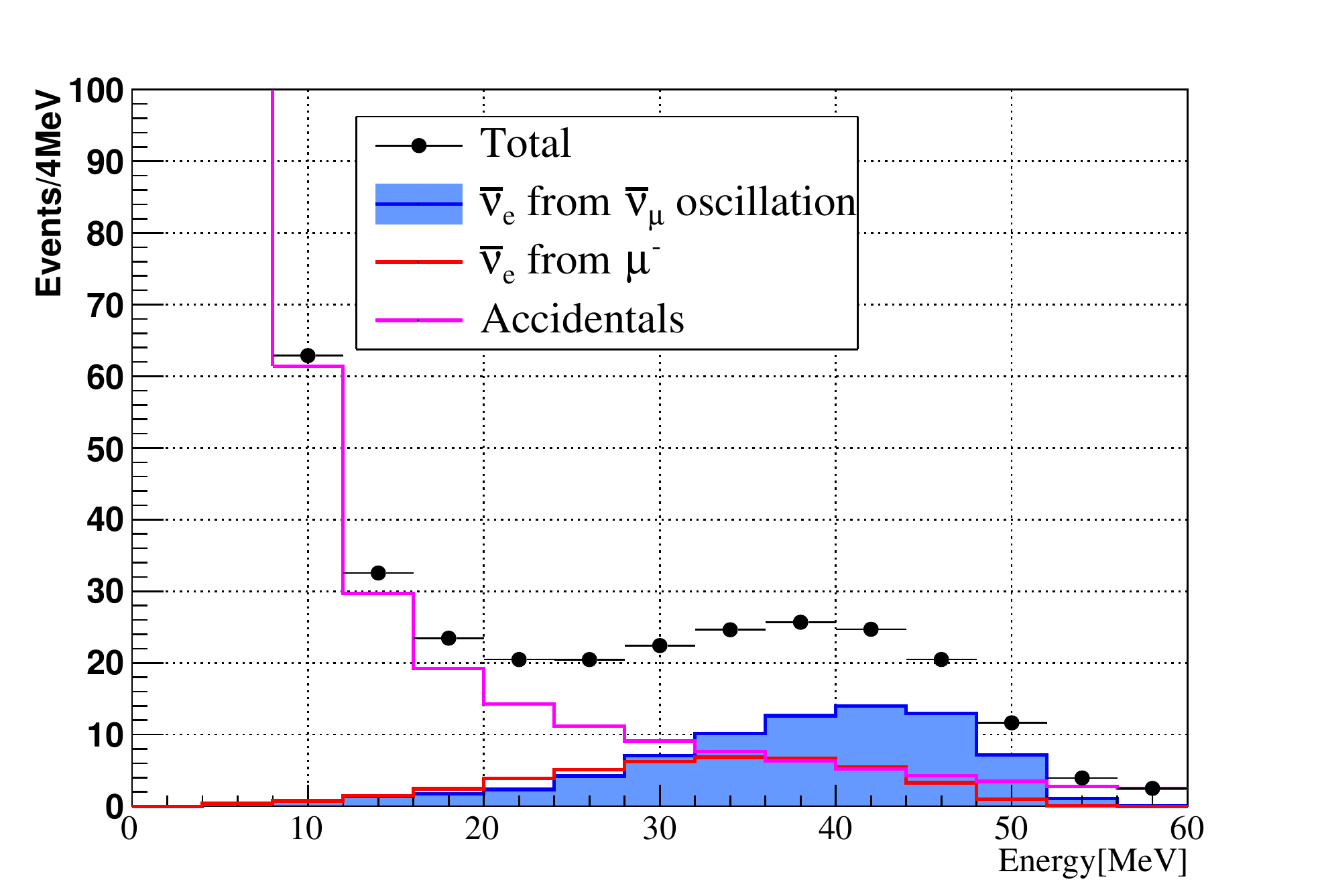}
}
\caption{\setlength{\baselineskip}{4mm}
  Expected observed energy spectrum in the JSNS$^2$-II in two detectors in the LSND best fit case for the 24 m detector (left) and the 48 m detector (right). 
} 
\label{FIG:WSterile}
\end{figure}

The expected energy spectra with the LSND best fit parameters of
neutrino oscillations ($\Delta m^2, sin^2 (2\theta)$)=(1.2, 0.003))
are shown in Fig.~\ref{FIG:WSterile}, 
The energy spectrum of the $\overline{\nu}_e$ generated by the
oscillation at 48m baseline is significantly different from
that of the intrinsic $\overline{\nu}_e$ background.

%%%%%%%%%%%%%%%%%%%%%%%%%%%%%%%%%%%%%%%%%%%%%%%%%%
\section{Fit methods and uncertainties}
\indent
%%%%%%%%%%%%%%%%%%%%%%%%%%%%%%%%%%%%%%%%%%%%%%%%%%

The important difference between the JSNS$^2$~\cite{CITE:TDR} and the
JSNS$^2$-II 
is the ability of JSNS$^2$-II to use different energy spectra
information with two different baselines.

\begin{figure}[htbp]
\centering
\subfigure[The energy spectrum of 24m detector]{
\includegraphics[width=0.45\textwidth,angle=0]{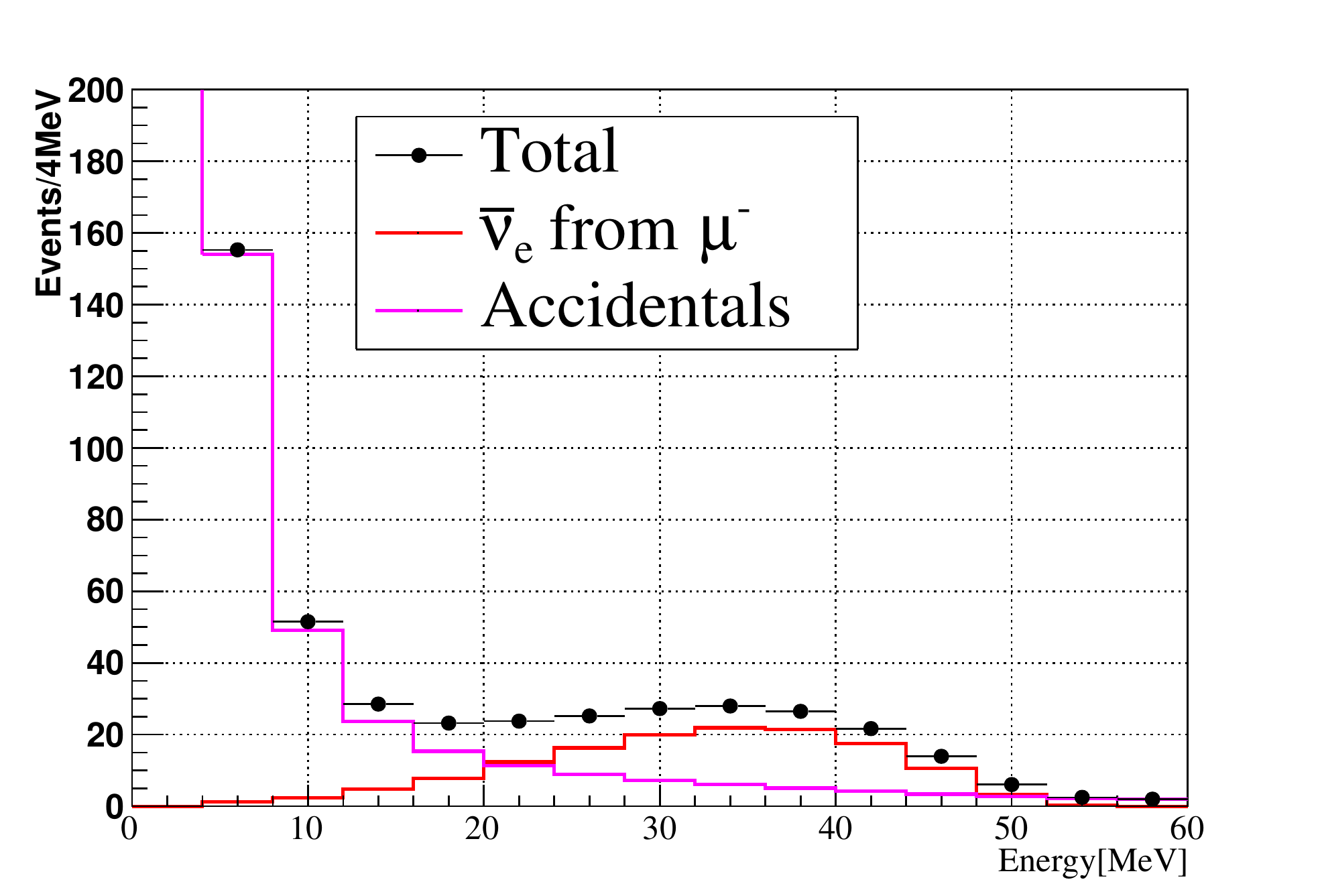}
}
\subfigure[The energy spectrum of 48m detector]{
\includegraphics[width=0.45\textwidth,angle=0]{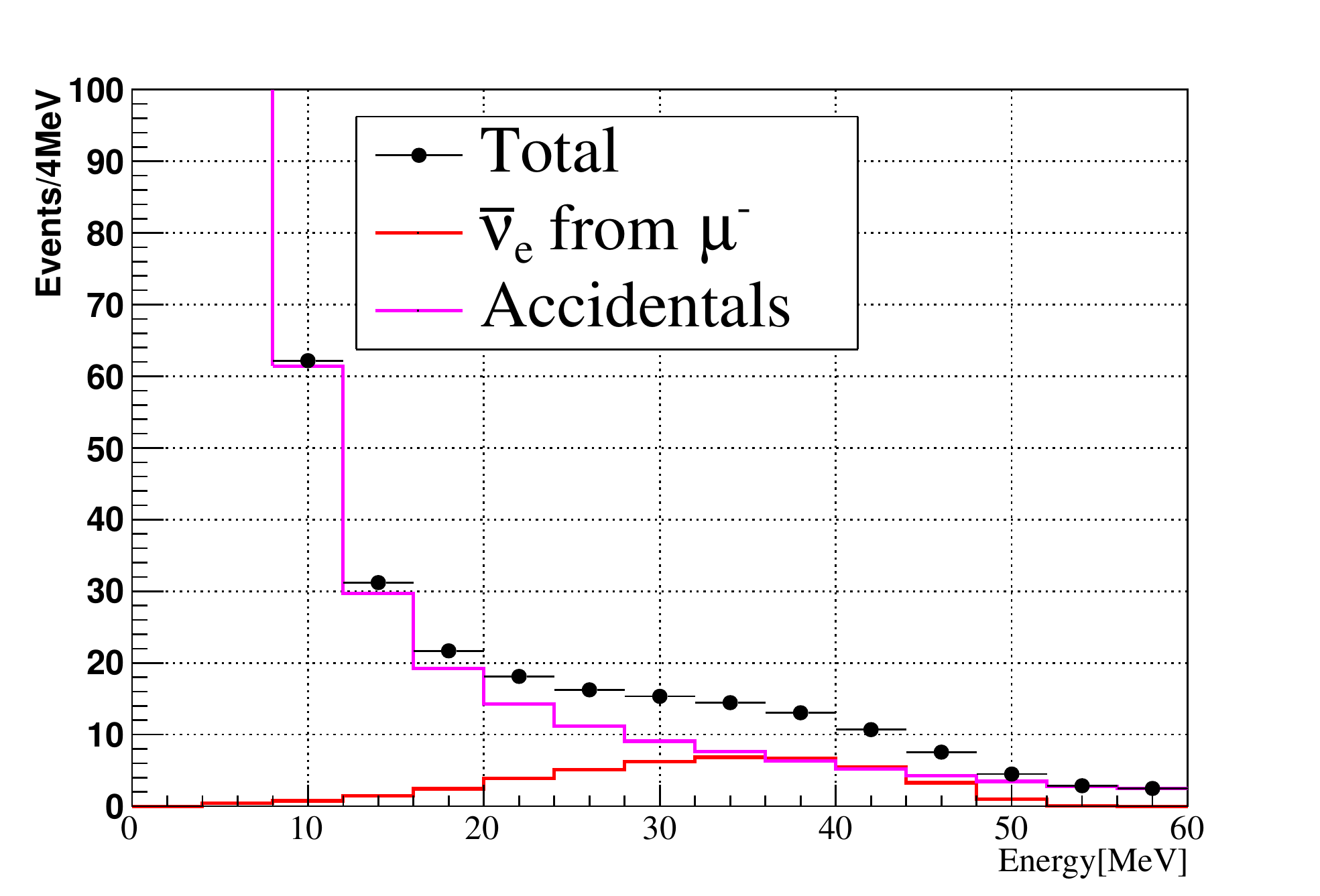}
}
\caption{\setlength{\baselineskip}{4mm}
  Expected observed energy spectrum in the JSNS$^2$-II in two detectors: 24 m detector (left), and 48 m detector (right).
} 
\label{FIG:NoSterile}
\end{figure}

Figure~\ref{FIG:NoSterile} shows the expected energy spectra in 24 m
detector and 48 m detector for the case  
with no neutrino oscillations in the JSNS$^2$-II.
To make the sensitivity plots, this pseudo-data is used.

As mentioned above, we have two background components.
One is from the accidental backgrounds, and
the other is from the intrinsic $\bar{\nu}_{e}$ from $\mu^{-}$.

The fitter considers what is the best mixture of the three components 
among (a) oscillated signal,
(b) accidental background, (c) intrinsic $\bar{\nu}_{e}$ from $\mu^{-}$
to explain the energy spectra in Fig.~\ref{FIG:NoSterile}.
Of course, it is crucial how accurately the normalization factors of these
three components are known, therefore the systematic uncertainties of the
factors will be discussed later.

%%%%%%%%%%%%%%%%%%%%%%%%%%%%%%%%%%%%%%%%%%%%%%%%%%
\subsection{Normalization uncertainty}
\indent
%%%%%%%%%%%%%%%%%%%%%%%%%%%%%%%%%%%%%%%%%%%%%%%%%%

For the fit of the oscillation parameters, $\Delta m^{2}$ and
$\sin^2(2\theta)$, constraints of the background normalization are 
important. However, $\bar{\nu}_{e}$ from $\mu^{-}$ has a
very poor normalization constraint from the external information 
since the production rates of charged pions are not well known
due that there have never been any experiments to study
the interactions between 3 GeV protons and mercury target. 
Therefore, 50$\%$ of the uncertainty of the normalization factor
for this background is used.
 
On the other hand, the cross section for
the $\nu_e + ^{12}C \rightarrow e + ^{12}N_{gs} $ reaction is
known at a 10$\%$ level~\cite{cite:XSEC12C}. The 
lifetime of $N_{gs} \beta$ decay and the $e^{-}$ energy spectrum are also 
well known. The measurement of the reaction provides the 
normalization factor for the oscillated signal
($\bar{\nu_{e}} + p \rightarrow e^{+} + n$) since the parent particle
for the oscillated signal is 
$\bar{\nu_{\mu}}$ from $\mu^{+}$ decays
($\mu^{+} \rightarrow e^{+} + \bar{\nu_{\mu}} + \nu_{e}$).
Note that the determination of the normalization factor can be done at the
10$\%$ level.

Finally, the normalization factor of the accidental background can
be very well determined by data since the main background of the IBD prompt
region is gamma rays induced by the cosmic rays and the IBD delayed
background is caused from the beam related gamma rays from the floor and
gamma rays induced by the cosmic rays
which have small time structure of the rate.
So, we don't need the specific systematic uncertainty   
on this background normalization.

%%%%%%%%%%%%%%%%%%%%%%%%%%%%%%%%%%
\subsection{Fit method}
%%%%%%%%%%%%%%%%%%%%%%%%%%%%%%%%%%
\indent

To get the exclusion region, we use the sum of the backgrounds
(intrinsic $\bar{\nu}_{e}$ and accidental backgrounds) as the
pseudo-data as mentioned above.
In this scheme, the maximum likelihood always obtains
$\sin^{2} 2 \theta$ = 0, regardless of the $\Delta m^2$.

%%%%%%%%%%%%%%%%%%%%%%%%%%%%%%%%%%
\subsubsection{Binned likelihood method}
%%%%%%%%%%%%%%%%%%%%%%%%%%%%%%%%%%
\indent

A binned maximum likelihood method is used for the analysis. The 
method fully utilizes the energy spectrum of each background and 
signal components, thus the amount of the signal components can be 
estimated efficiently.

For this purpose, the following equation is used.
\begin{eqnarray}
\label{Eq:likelihood}
L &=& L_{1} \times L_{2} \\ 
&=& \displaystyle \left[ \Pi_{i} P_{1}(N_{exp} | N_{obs})_{i} \right]_{1} \times \left[ \Pi_{i} P_{2}(N_{exp} | N_{obs})_{i} \right]_{2}\\
P(N_{exp} | N_{obs}) &=& \frac{e^{-N_{exp}} \cdot (N_{exp})^{N_{obs}}}{N_{obs}!}
\end{eqnarray}
where, $L_{1}$ and $L_{2}$ are the likelihoods of the first and second detectors,
the index $i$ refers to the $i$-th energy bin, $N_{exp}$ is expected 
number of events in $i$-th bin, $N_{obs}$ is number of observed
events in $i$-th bin.
%$i$ is starting from 20 MeV and ends at 60 MeV
Energies between 20 and 60 MeV are only considered 
because the energy cut above 20 MeV is applied for the primary signal
as explained before. Note that
$N_{exp} =  N_{sig}(\Delta m^{2}, \sin^{2}2\theta) + \displaystyle\sum N_{bkg}$,
and $N_{sig}(\Delta m^{2}, \sin^{2}2\theta)$ is calculated 
by the two flavor neutrino oscillation equation as shown before: 
$P(\bar{\nu_{\mu}} \rightarrow \bar{\nu_{e}}) =
\sin^{2}2\theta \sin^{2}(\frac{1.27 \cdot \Delta m^{2} (eV^{2})
\cdot L (m)}{E_{\nu} (MeV)})$.

The maximum likelihood point gives the best fit parameters, and
2$\Delta lnL$ provides the uncertainty of the fit parameters. As shown in the 
PDG~\cite{CITE:PDG}, we have
to use the 2$\Delta lnL$ for 2 parameter fits to determine the
uncertainties from the fit.   
 
%%%%%%%%%%%%%%%%%%%%%%%%%%%%%%%%%%
\subsubsection{Treatment of systematic uncertainties}
%%%%%%%%%%%%%%%%%%%%%%%%%%%%%%%%%%
\indent

Equation~\ref{Eq:likelihood} takes only statistical 
uncertainty into account, therefore the systematic uncertainties
should be incorporated in the likelihood.  
Fortunately, energy spectrum of the oscillated signal and background
components are well known, thus the error (covariance) matrix of 
energy is not needed. Note that the uncertainty on the energy scale
does not affect to the sensitivity
if we achieve the 1$\%$ systematic error as described
in \cite{CITE:TDR2} (See Fig.121 therein).

In this case, the uncertainties of the overall normalization
of each component have to be taken into account, and the assumption
is a good approximation. 

In order to incorporate the systematic uncertainties, the constraint terms
should be added to Equation~\ref{Eq:likelihood} and the
equation is changed as follows.

\begin{eqnarray}
L &=& [\Pi_{i} P( N_{exp}^{'} | N_{obs})_{i}] \times e^{- \frac{(1-f_{1})^{2}}{2 \Delta \sigma_{1}^{2}}} \times e^{- \frac{(1-f_{2})^{2}}{2 \Delta \sigma_{2}^{2}}}  
\label{Eyq:likelihood2}
\end{eqnarray}\
where $f_{j}$ are nuisance parameters to give the constraint term on the
overall normalization factors.
$N_{exp}^{'} = f_{1} \cdot N_{sig} (\Delta m^{2}, \sin^{2}2\theta)
+ f_{2} \cdot N_{intrinsic \hspace*{0.1cm} bkg} + N_{accidental \hspace*{0.1cm} bkg}$.
$\Delta \sigma_{i}$
gives the uncertainties
on the normalization factors of each components.
In this proposal, the profiling fitting method is used to treat the systematic 
uncertainties. The method is widely known as the correct fitting method as
well as the marginalizing method.
The profiling method fits all nuisance parameters as well as 
oscillation parameters. 

As mentioned above, 
the flux of the $\bar{\nu}_{e}$ from $\mu^{-}$ decays around 
the mercury target has very poor constraints from the external information.
For this situation, the uncertainty of this background
component is assigned to be 50$\%$.
%Note again, we assume
%NA61 efforts makes this uncertainty to be 10$\%$ in later sections also.

Table~\ref{tab:constraints} shows the summary of the 
uncertainty of the normalization factors for the signal and background 
components. They are regarded as inputs of $\Delta \sigma$ although only 
$\bar{\nu}_{e}$ from $\mu^{-}$ and the accidental background are
used as mentioned above.

\begin{table}[h]
\begin{center}
	\begin{tabular}{|l|c|c|}
	  \hline
	  components  & uncertainty & comments \\ \hline \hline
	  signal      & 10$\%$ & normalized by $\nu_e$ from $\mu^{+}$ \\ \hline
	  $\bar{\nu_{e}}$ from $\mu^{-}$  & 50  $\%$ &   \\ \hline
	  cosmic / beam & negligible & well known from calibration source \\ \hline
	\end{tabular}
	\caption{\setlength{\baselineskip}{4mm}
	  Summary of uncertainties on the normalization factors. Note that only $\bar{\nu}_{e}$ from $\mu^{-}$ and accidental backgrounds are used in the fitting since they are dominant ones in TDR.}
	\label{tab:constraints}
\end{center}
\end{table}

%%%%%%%%%%%%%%%%%%%%%%%%%%%%%%%%%%%%%%%%%%%%%%%%%%
\section{Expected sensitivity}  
\label{Sec:ES}
\indent
%%%%%%%%%%%%%%%%%%%%%%%%%%%%%%%%%%%%%%%%%%%%%%%%%%

\begin{figure}[htbp]
  \centering
  \subfigure{
    \includegraphics[width=0.4\textwidth,angle=0]{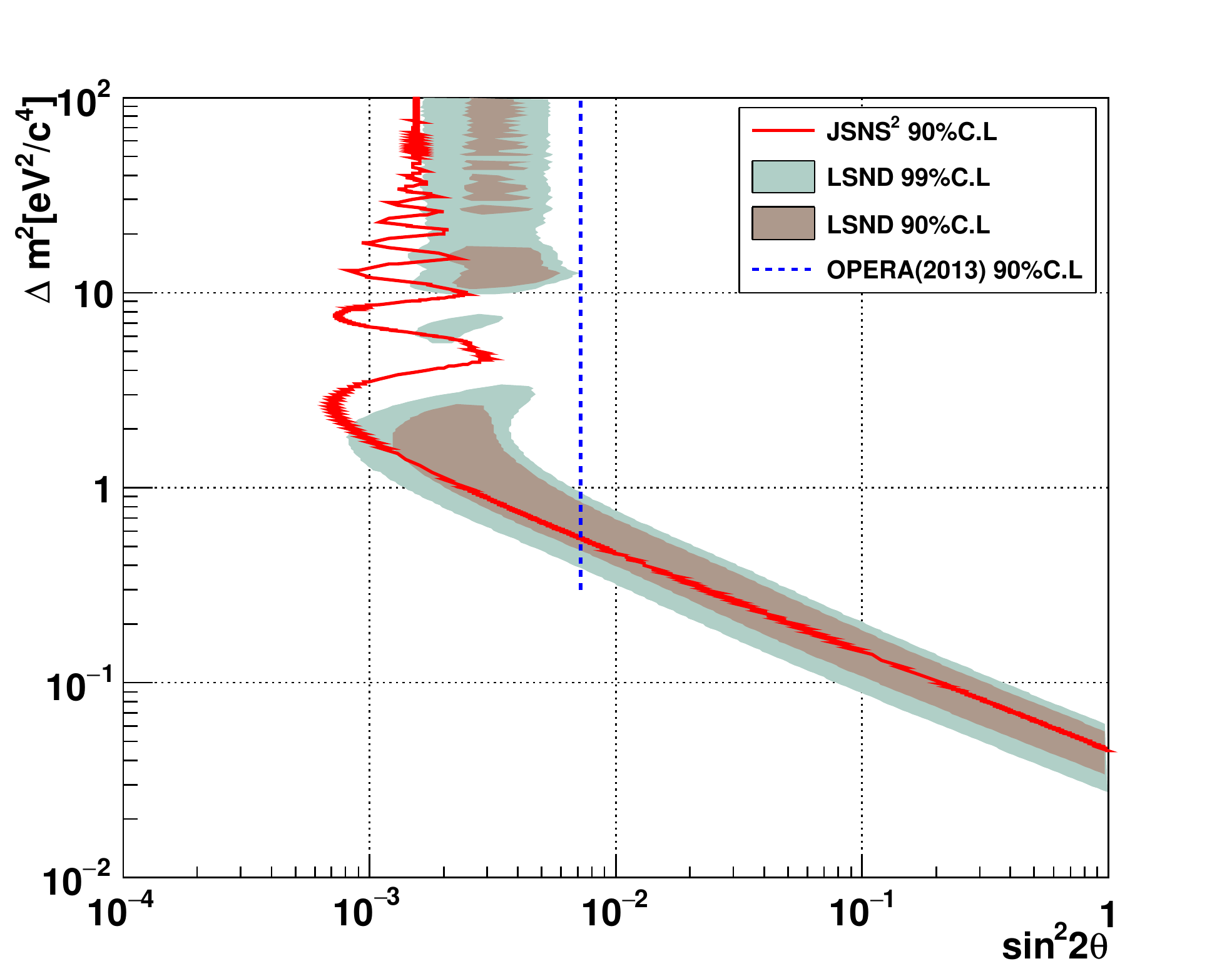}
  }
  \subfigure{
    \includegraphics[width=0.55\textwidth,angle=0]{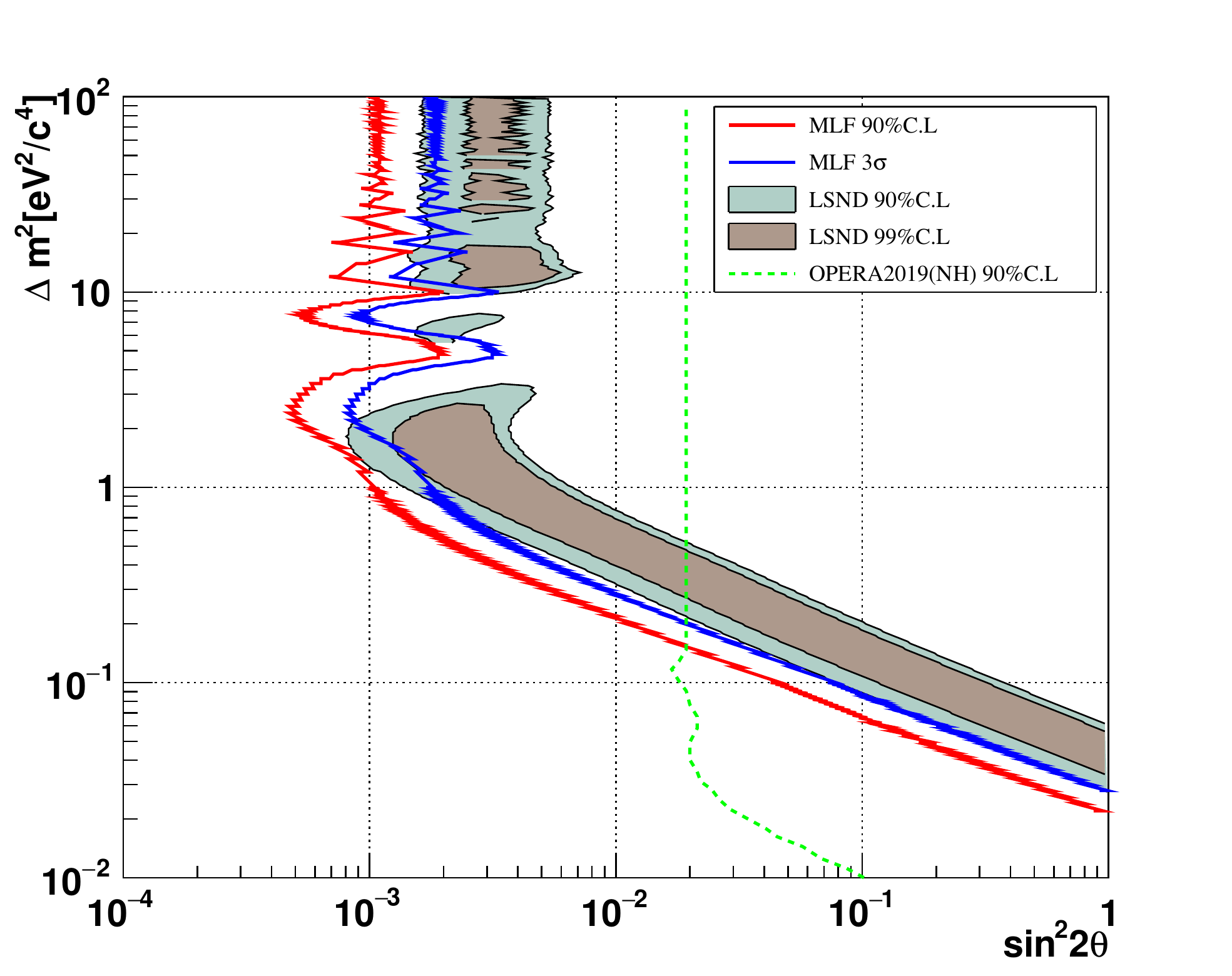}
  }
    \caption{\setlength{\baselineskip}{4mm}
      The sensitivity of the JSNS$^2$-II (right). For the comparison,
      the sensitivity of the current JSNS$^2$ is also shown in the left.
    } 
\label{Fig:2448}
\end{figure}

Figure~\ref{Fig:2448} shows the sensitivity of the JSNS$^2$-II.
We assume that the starting point of the JSNS$^2$-II is after 3 years
of running of the current JSNS$^2$ experiment. For the reference,
the sensitivity of the current JSNS$^2$ experiment~\cite{CITE:TDR}
is also shown.

The sensitivity becomes better, especially in the low $\Delta m^2$ region,
which has been indicated by the global fit of the appearance
mode~\cite{CITE:Review} because of the longer baseline. The
3$\sigma$ C.L. line nicely covers most of LSND indicated regions.
%and the LSND best fit point is concluded in 5$\sigma$ C.L.

\begin{figure}[htbp]
\centering
\subfigure[near (24 m)]{
\includegraphics[width=0.3\textwidth,angle=0]{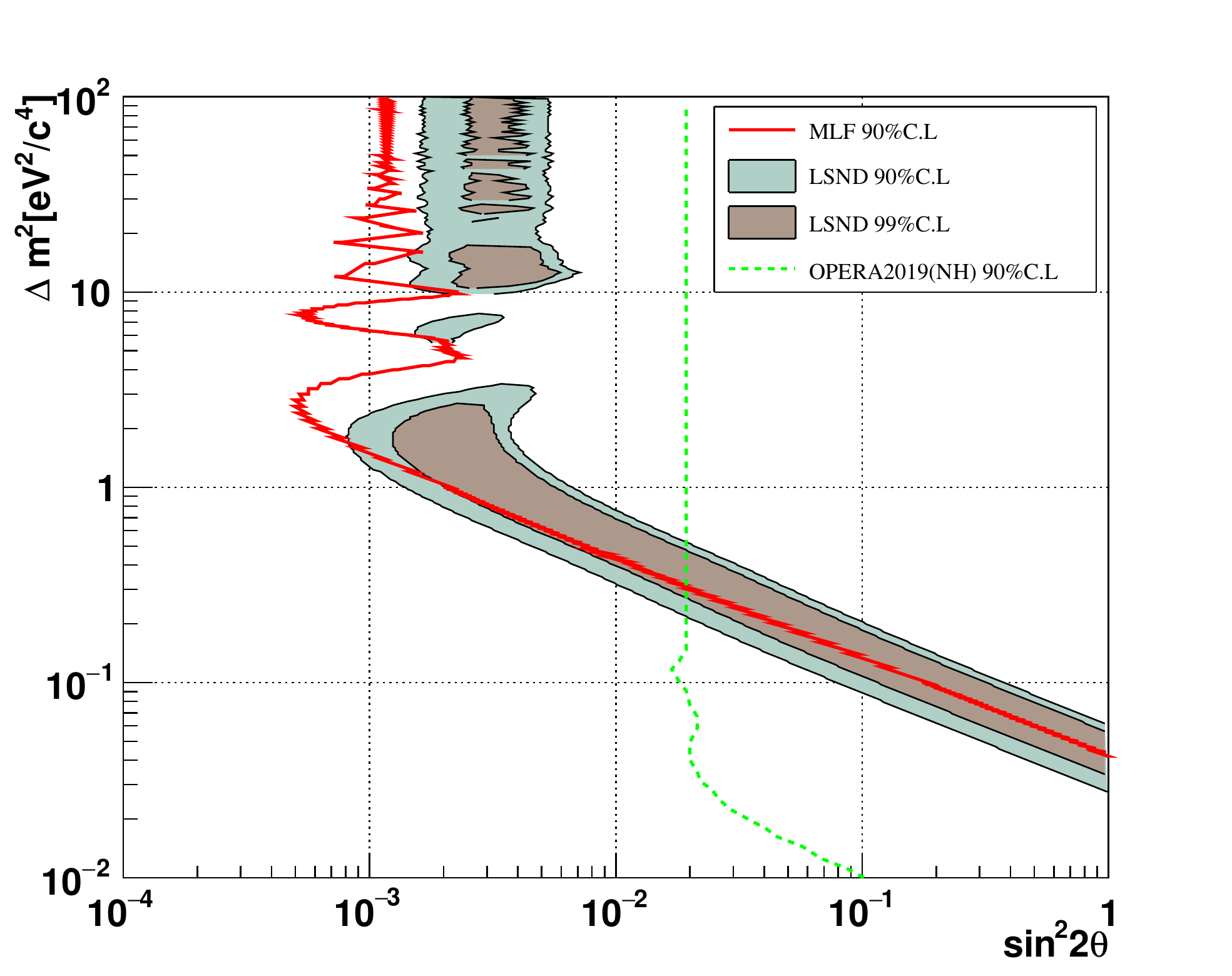}
}
\subfigure[far (48 m)]{
\includegraphics[width=0.31\textwidth,angle=0]{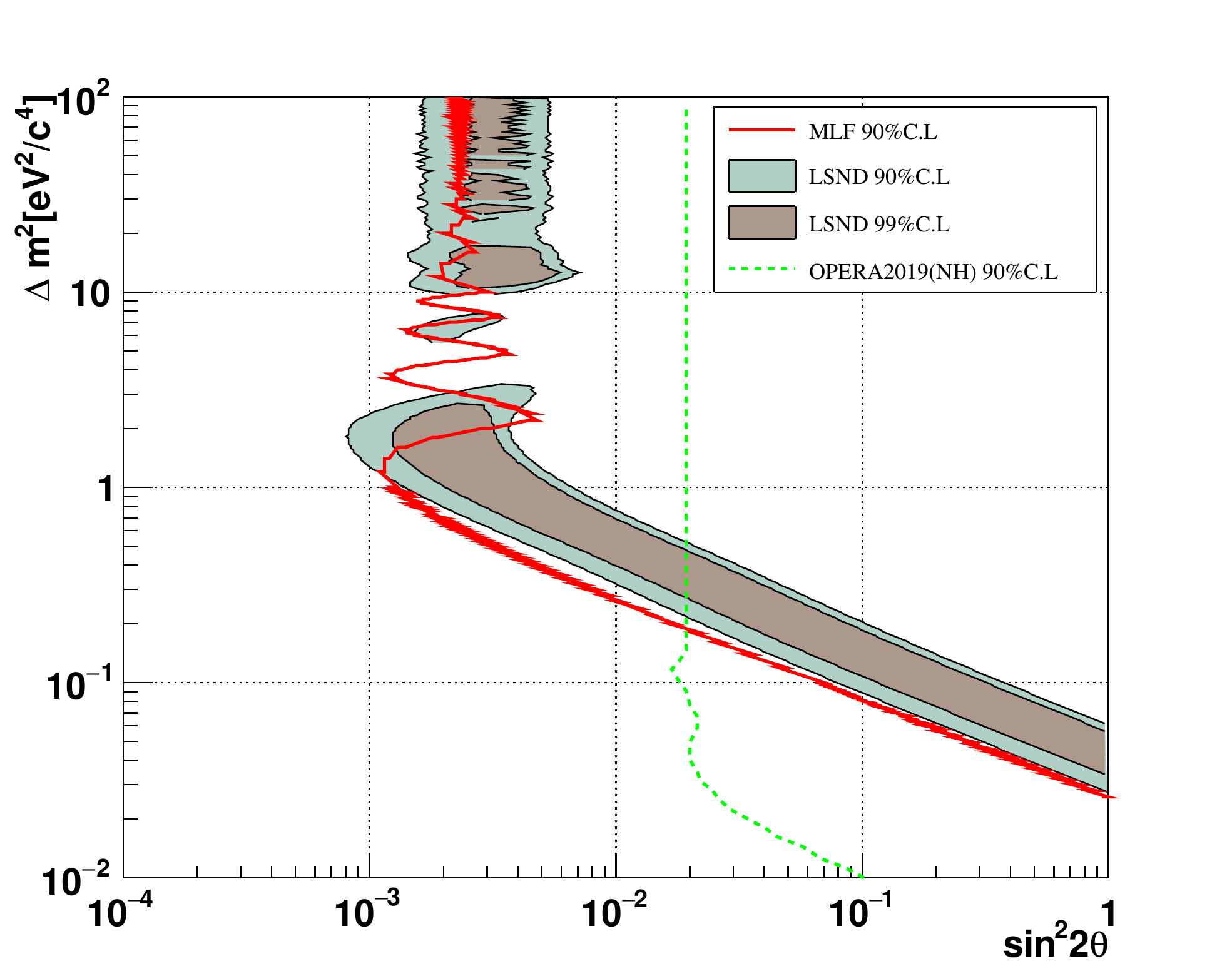}
}
\subfigure[each detector/combined]{
\includegraphics[width=0.31\textwidth,angle=0]{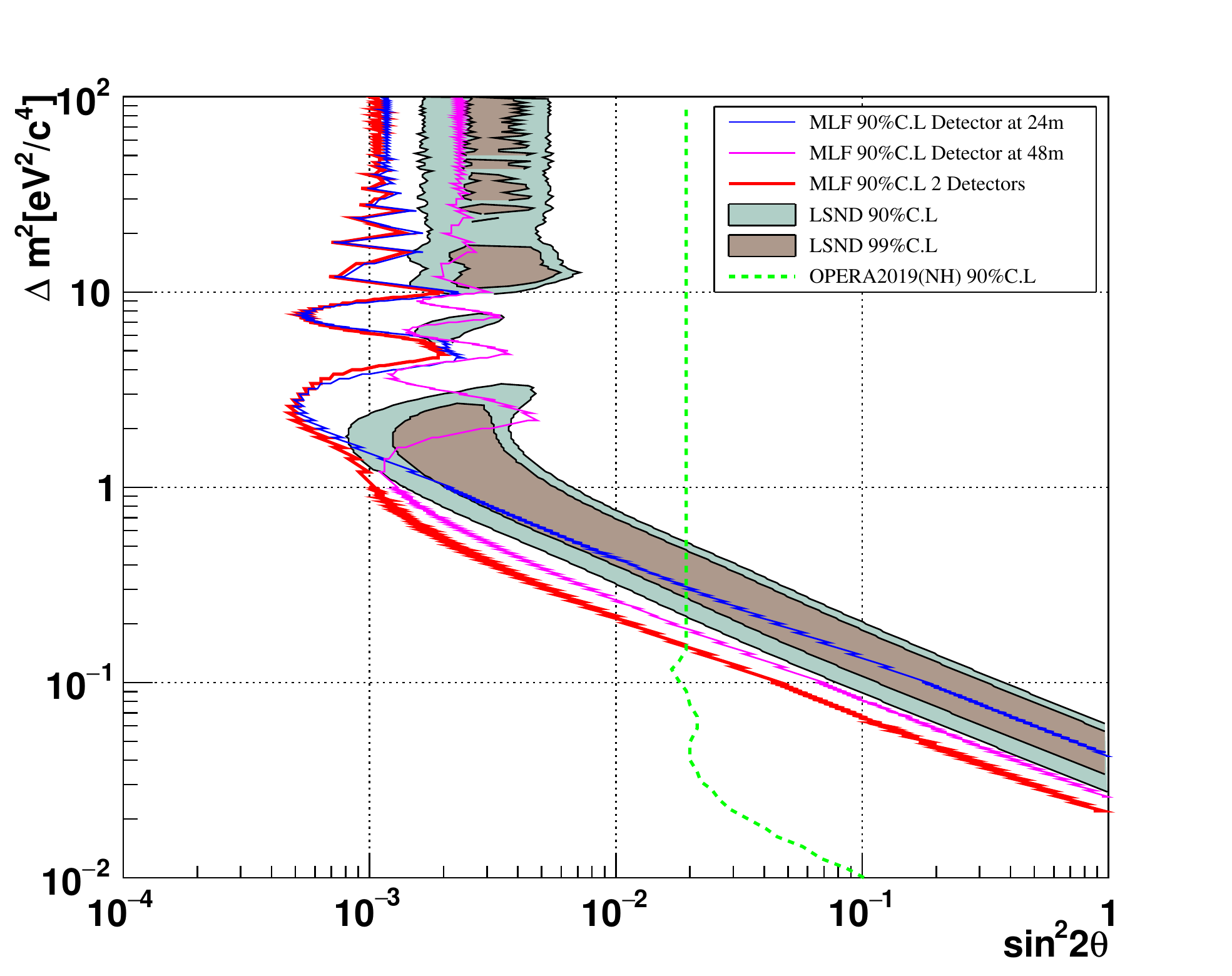}
}
\caption{\setlength{\baselineskip}{4mm}
The sensitivity of the near and the far detectors. (a): the near (24 m) detector only. (b): the far (48 m) detector only. (c): overlaid plots for each detctor and the JSNS$^2$-II.  
} 
\label{Fig:24482}
\end{figure}

To understand the power of the sensitivity of each detector
(24 m and 48 m baselines) and
combination of the signal and background understanding,
Figs.~\ref{Fig:24482} compare the sensitivity with each detector only and
JSNS$^2$-II.
Below 1 eV$^2$, the second detector effects are large as expected.
Also we can see the cancellation effects of the systematic uncertainties
on the neutrino flux normalizations, espaecially at lower $\Delta m^2$.
In the lower $\Delta m^2$ region,
the energy spectra between the oscillation signals
and the intrinsic background are getting closer due to the oscillation pattern.
In that case, the cancellation effects of the normalization error on
the intrinsic background due to two detectors are large.

%%%%%%%%%%%%%%%%%%%%%%%%%%%%%%%%%%%%%%%%%%%%%%%%%%
\section{Timescale and cost}  
\indent
%%%%%%%%%%%%%%%%%%%%%%%%%%%%%%%%%%%%%%%%%%%%%%%%%%

\begin{figure}[htbp]
  \begin{center}
  \includegraphics[width=0.7\textwidth,angle=0]{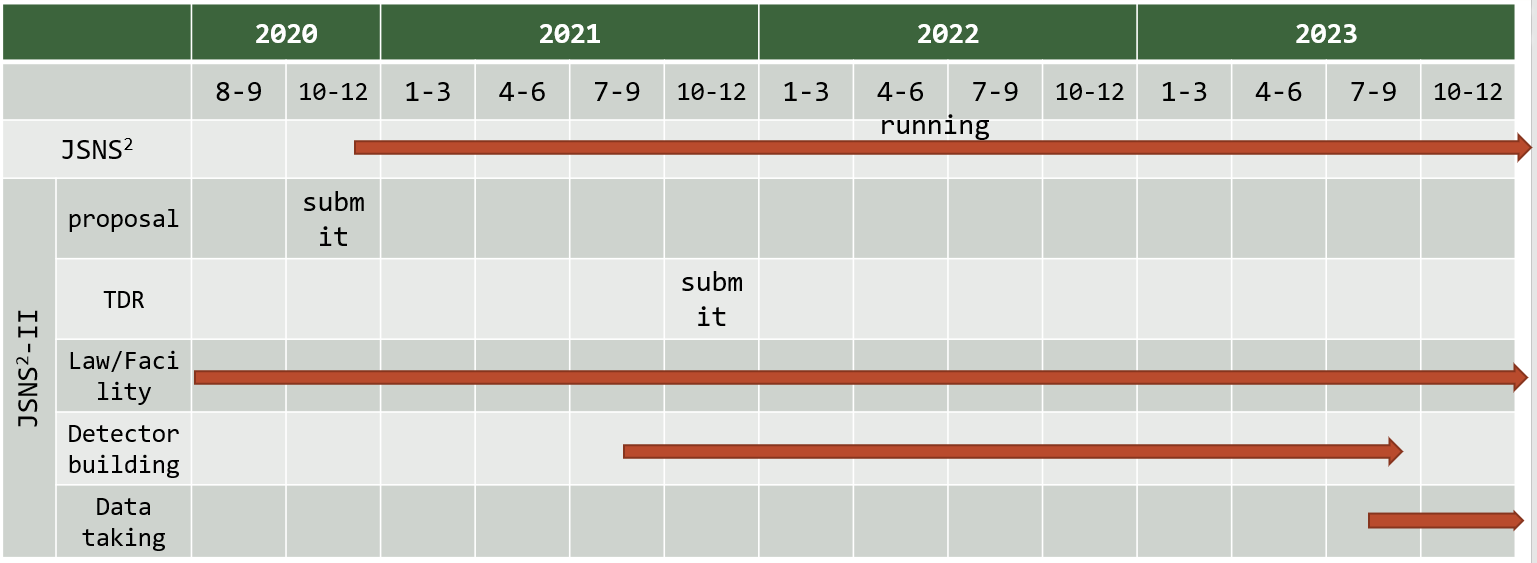}
  \caption{\setlength{\baselineskip}{4mm}
    The timescale of the JSNS$^2$-II. }
  \label{Fig:Timescale}
  \end{center}
\end{figure}

The possible timescale of the JSNS$^2$-II is shown in Fig.~\ref{Fig:Timescale}.
Whilst running the current JSNS$^2$ experiment,
the second detector will be built, and the JSNS$^2$-II will start
data taking from FY2023.

Table~\ref{Tab:Cost} shows the cost estimation of the second detector.
These numbers are based on the experience of the 1st detector construction.
Note that the Grant-in-Aid for Specially Promoted Research in Japan
is adpoted in FY2020, therefore 480M Yen was granted to this project.
We assume some of components could be donated from the foreign experiment(s)
here.
\begin{table}[htbp]
  \begin{center}
   \begin{tabular}{lrrr} 
    \hline
  \hline
  Item                             &Unit price  & Quantity  & Total    \\
    \hline
* PMTs \& Electronics system :        & 500Ky/ch  & 240~ch  & 50My      \\
* Tanks \& Acrylic Vessels :          & 120My/set  &  & 120My     \\
* Gd-LS, Buffer-LS	                  &           &        & 10My  \\
* Fluid handling and infrastructure   & 50My/set  & 1~set  &  50My     \\
* Miscellaneous                       &           &        &  50My   \\
\hline
\hline
Grand Total	                     &            &        & 280My \\
   \end{tabular}
   \caption{Cost estimation (the second (far) detector). Donations from
   other foreign experiment(s) are assumed.}
   \label{Tab:Cost}
 \end{center}
\end{table}

%%%%%%%%%%%%%%%%%%%%%%%%%%%%%%%%%%%%%%%%%%%%%%%%%%
\section{Acknowledements}  
\indent
%%%%%%%%%%%%%%%%%%%%%%%%%%%%%%%%%%%%%%%%%%%%%%%%%%

We thank the J-PARC staff for their support. We acknowledge the support of the Ministry of Education, Culture, Sports, Science, and Technology (MEXT) and the JSPS grants-in-aid (Grant Number 16H06344, 16H03967, 20H05624), Japan. This work is also supported by the National Research Foundation of Korea (NRF) Grant No. 2016R1A5A1004684, 2017K1A3A7A09015973, 2017K1A3A7A09016426, 2019R1A2C3004955, 2016R1D1A3B02010606, 2017R1A2B4011200, 2018R1D1A1B07050425, 2020K1A3A7A0908
0133 and 2020K1A3A7A09080114. Our work has also been supported by a fund from the BK21 of the NRF. The University of Michigan gratefully acknowledges the support of the Heising-Simons Foundation. This work conducted at Brookhaven National Laboratory was supported by the U.S. Department of Energy under Contract DE-AC02-98CH10886. The work of the University of Sussex is supported by the Royal Society grant no.IES\textbackslash R3\textbackslash 170385

%%%%%%%%%%%%%%%%%%%%%%%%%%%%%%%%%%%%%%%%%%%%%%%%%%
% bibliography
%%%%%%%%%%%%%%%%%%%%%%%%%%%%%%%%%%%%%%%%%%%%%%%%%%

\end{document}